\documentclass[graybox]{svmult}

\usepackage{amsmath,amssymb}
\usepackage{mathptmx}       
\usepackage{helvet}         
\usepackage{courier}        
\usepackage{type1cm}        
\usepackage{makeidx}         
\usepackage{graphicx}        
\usepackage{multicol}        
\usepackage[bottom]{footmisc}

\makeindex             

\DeclareMathOperator{\extdm}{d}
\newcommand{\extd}{\extdm \!}

\begin{document}

\title*{Black holes and thermodynamics --- The first half century}
\titlerunning{Black hole thermodynamics}
\author{Daniel Grumiller, Robert McNees and Jakob Salzer}
\institute{Daniel Grumiller \at Institute for Theoretical Physics, Vienna University of Technology, Wiedner Hauptstrasse 8-10, A-1040 Vienna, Austria \email{grumil@hep.itp.tuwien.ac.at}
\and Robert McNees \at Loyola University Chicago, Department of Physics, Chicago, IL 60660, USA \email{rmcnees@luc.edu}
\and Jakob Salzer \at Institute for Theoretical Physics, Vienna University of Technology, Wiedner Hauptstrasse 8-10, A-1040 Vienna, Austria \email{salzer@hep.itp.tuwien.ac.at}}
%
%
\maketitle

\abstract*{Black hole thermodynamics emerged from the classical general relativistic laws of black hole mechanics, summarized by Bardeen-Carter-Hawking, together with the physical insights by Bekenstein about black hole entropy and the semi-classical derivation by Hawking of black hole evaporation. The black hole entropy law
inspired the formulation of the holographic principle by 't Hooft and Susskind, which is famously realized in
the gauge/gravity correspondence by Maldacena, Gubser-Klebanov-Polaykov and Witten within string theory.
Moreover, the microscopic derivation of black hole entropy, pioneered by Strominger-Vafa within string theory,
often serves as a consistency check for putative theories of quantum gravity. In this book chapter we review
these developments over five decades, starting in the 1960ies.}

\abstract{Black hole thermodynamics emerged from the classical general relativistic laws of black hole mechanics, summarized by Bardeen--Carter--Hawking, together with the physical insights by Bekenstein about black hole entropy and the semi-classical derivation by Hawking of black hole evaporation. The black hole entropy law
inspired the formulation of the holographic principle by 't Hooft and Susskind, which is famously realized in the gauge/gravity correspondence by Maldacena, Gubser--Klebanov--Polaykov and Witten within string theory. Moreover, the microscopic derivation of black hole entropy, pioneered by Strominger--Vafa within string theory, often serves as a consistency check for putative theories of quantum gravity. In this book chapter we review these developments over five decades, starting in the 1960ies.}
 
\section*{Introduction and Prehistory}


\runinhead{Introductory remarks.}
The history of black hole thermodynamics is intertwined with the history of quantum gravity. 
In the absence of experimental data capable of probing Planck scale physics the best we can do is to subject putative theories of quantum gravity to stringent consistency checks. Black hole thermodynamics provides a number of highly non-trivial consistency checks. Perhaps most famously, any theory of quantum gravity that fails to reproduce the Bekenstein--Hawking relation
\begin{equation}
 S_{\textrm{\tiny BH}} = \frac{k_{B}\,c^{3} A_h}{4\hbar G}
\label{eq:td1}
\end{equation}
between the black hole entropy $S_{\textrm{\tiny BH}}$, the area of the event horizon $A_h$, and Newton's constant $G$ would be regarded with a great amount of skepticism (see e.g~\cite{Carlip:2001wq}).

In addition to providing a template for the falsification of speculative models of quantum gravity, black hole thermodynamics has also sparked essential developments in the field of quantum gravity and remains a vital source of insight and new ideas. Discussions about information loss, the holographic principle, the microscopic origin of black hole entropy, gravity as an emergent phenomenon, and the more recent firewall paradox all have roots in black hole thermodynamics. Furthermore, it is an interesting subject in its own right, with unusual behavior of specific heat, a rich phenomenology, and remarkable phase transitions between different spacetimes.


In this review we summarize the development of black hole thermodynamics chronologically, except when the narrative demands deviations from a strictly historical account. While we have tried to be comprehensive, our coverage is limited by a number of factors, not the least of which is our own knowledge of the literature on the subject. Each of the following five sections describes a decade, beginning with the discovery of the Kerr solution in 1963 \cite{Kerr:1963ud}.
In our concluding section we 
look forward to future developments. But before starting we comment on some early insights that had the potential to impact the way we view the result \eqref{eq:td1}.

\runinhead{Prehistory.}
  If the history of black hole thermodynamics begins with the papers of Bekenstein \cite{Bekenstein:1972tm} and Bardeen, Carter, and Hawking \cite{Bardeen:1973gs}, then the prehistory of the subject stretches back nearly forty additional years to the work of Tolman, Oppenheimer, and Volkoff in the 1930s \cite{tolman1987relativity,Oppenheimer:1939ne,Tolman:Static}. These authors considered the conditions for a `star' -- a spherically symmetric, self-gravitating object composed of a perfect fluid with a linear equation of state -- to be in hydrostatic equilibrium. 
Later, in the 1960s, Zel'dovich showed that linear equations of state besides the familiar $p=0$ (dust) and $p=\rho/3$ (radiation) are consistent with relativity \cite{Zel:1961}. He established the bound $p \leq \rho$, with $p = \rho$ representing a causal limit where the fluid's speed of sound is equal to the speed of light. A few years after that, Bondi considered massive spheres composed of such fluids and included the case $p = \rho$ in his analysis \cite{Bondi:Massive}.

  The self-gravitating, spherically symmetric perfect fluids considered by these and other authors possess interesting thermodynamic properties. In particular, the entropy of such objects (which are always outside their Schwarzschild radius) is not extensive in the usual sense. For example, a configuration composed of radiation has an entropy that scales with the size of the system as $S(R) \sim R^{3/2}$, and a configuration with the ultra-relativistic equation of state $p = \rho$ has an entropy $S(R) \sim R^2$ that scales like the area. But these results do not appear in the early literature (at least, not prominently) because there was no compelling reason to scrutinize the relationship between the entropy and size of a gravitating system before the 1970s. It was not until the 1980s, well after the initial work of Bekenstein and Hawking, that  Wald, Sorkin, and Zhang studied the entropy of self-gravitating perfect fluids with $p=\rho/3$ \cite{Sorkin:1981wd}. They showed that the conditions for hydrostatic equilibrium -- the same conditions set out by Tolman, Oppenheimer, and Volkoff -- give at least local extrema of the entropy. With reasonable physical assumptions these objects quite easily satisfy the Bekenstein bound, $S\leq 2\pi k_B R E/(\hbar c)$, where $R$ and $E$ are the object's size and energy, respectively.

  The area law \eqref{eq:td1} is often presented as a surprising deviation from the volume scaling of the entropy in a non-gravitating system. But the early work described above suggests, without invoking anything as extreme as a black hole, that this is something we should expect from General Relativity. Even a somewhat mundane system like a sufficiently massive ball of radiation has an entropy that is not proportional to its volume. The surprising thing about the area law is not that the entropy of the system grows much more slowly than a volume. Rather, it is that the entropy of a black hole seems to saturate, at least parametrically, an \emph{upper} bound on the growth of entropy with the size of a gravitating system. Such a bound, which follows from causality, could have been conjectured several years before the work of Bekenstein and Hawking.

  



\section*{1963--1973}
 

\runinhead{Black hole solutions and the uniqueness theorem.}
After the first black hole solutions were found in immediate consequence to the publication of Einstein's equations, it took almost 50 years for the next exact black hole solution to be discovered. The Kerr solution \cite{Kerr:1963ud} describes a rotating black hole of mass $M$ and angular momentum $J=aM$ 
\begin{multline}
\extd s^2=-\Big(1-\frac{2Mr}{\rho^2}\Big)\,\extd t^2-\frac{4Mra\sin^2\!\theta}{\rho^2} \,\extd t\,\extd\phi+\Big(r^2+a^2+\frac{2Mra^2\sin^2\!\theta}{\rho^2}\Big)\sin^2\!\theta\,\extd\phi^2\\
+\frac{\rho^2}{r^2-2Mr+a^2} \,\extd r^2+\rho^2\,\extd\theta^2 \qquad \qquad\textrm{with}\; \rho^2 := r^2+a^2\cos^2\!\theta\,.
\label{eq:kerr}
\end{multline}
Only two years later this solution was extended to include charged rotating black holes \cite{Newman:1965yu}. These black hole solutions exhibit the remarkable property that they are parameterized in terms of only three quantities as measured from infinity: mass, angular momentum, charge. 
It was therefore natural to ask whether this was the case for all black hole solutions.\par 
Building on earlier work concerning the persistence of the horizon under asymmetric perturbations \cite{Ginzburg:1965uf,Doroshkevich:1966ru}, Israel proved that --- assuming some regularity conditions --- the Schwarzschild solution is the only static, asymptotically flat vacuum spacetime that exhibits a regular horizon \cite{Israel:1967wq}. Later, this proof was generalized to static asymptotically flat electrovac spacetimes, now with the Reissner--Nordstr\"om black hole as the only admissible spacetime \cite{Israel:1967za}. In the case of axisymmetric stationary black holes Carter was later able to show that these spacetimes fall into discrete sets of continuous families, each of them depending on one or two independent parameters, with the Kerr solutions as the unique family to allow vanishing angular momentum \cite{Carter:1971zc}. The key point of Carter's proof is the observation that Einstein's equations for an axisymmetric spacetime can be reduced to a two-dimensional boundary value problem. Building on this, Robinson showed that in fact only the Kerr family exists, thus establishing the uniqueness of the Kerr black hole \cite{Robinson:1975bv}. Similar results concerning the classification and uniqueness of charged axisymmetric stationary black holes were worked out independently by Mazur \cite{Mazur:1982db}, Bunting \cite{Bunting:1983bh} and more recently by Chrusciel and Costa \cite{Chrusciel:2008js}. However, due to different initial hypotheses in the statement of the theorem and some technical gaps, the uniqueness theorem is still extensively studied (cf. \cite{Chrusciel:2012jk} for a review).\par
Referring to these results, John Wheeler coined the expression ``black holes have no hair'' \cite{Misner:1973qy}, i.e. black holes can be described entirely by a small amount of quantities measured from infinity. The no-hair conjecture thus suggests a resemblance of black holes to systems in thermodynamic equilibrium, whose macroscopic state is parameterized by a small number of macroscopic variables. \par

\runinhead{Penrose process and superradiant scattering.}
Another similarity between black holes and thermodynamical systems was revealed with Penrose's suggestion that energy can be extracted from a rotating black hole \cite{Penrose:1969pc}. The Penrose process relies on the presence of an ergosphere in Kerr spacetime. In this region the Killing field $\xi^{a}$ that asymptotically corresponds to time translation is spacelike. Consequently, the energy $E=-p_a \xi^{a}$ of a particle of 4-momentum $p^{a}$ need not be positive. In the Penrose process a particle with positive energy $E_0$ is released from infinity. In the ergosphere the particle breaks up in such a way that one fragment has negative energy $E_1$ whereas the other has positive energy $E_2=E_0-E_1>E_0$. If the latter returns back to infinity on a geodesic one has effectively gained the energy $|E_1|$. The negative energy particle falls into the black hole and therefore reduces its mass. Thus, energy is indeed extracted from the black hole. Angular momentum $j^{a}_2$ and energy of the particle falling into the black hole have to obey the inequality $j^{a} \le E_2/\Omega_H$, where $\Omega_H$ is the angular velocity of the black hole. Therefore, the change in the black hole's mass and angular momentum $\delta M$ and $\delta J$, respectively, are related by $\delta M\ge \Omega_H \delta J$. 
This equation can be rewritten in a form that bears a clear resemblance to the second law of thermodynamics \cite{Christodoulou:1970wf}
\begin{equation}
\delta M_{irr}\ge 0, \label{eq:done1}
\end{equation}
where $M^2_{irr}=\frac{1}{2}\left(M^2+\sqrt{M^4-J^2}\right)$ is the irreducible mass.
Expressed in terms of irreducible mass and angular momentum, the mass of the black hole reads
\begin{equation}
M^2=M_{irr}^2+\frac{J^2}{4M_{irr}^2}\ge M^2_{irr}.
\end{equation}
The maximum amount of energy that can be extracted from a black hole with initial mass $M_0$ and angular momentum $J_0$ is therefore $\Delta M=M_0-M_{irr}(M_0,J_0)$, which is maximized for an extremal black hole, i.e. $J_0=M_0^2$, with an efficiency of $0.29$. A generalization to charged rotating black holes yields the Christodoulou--Ruffini mass formula
\begin{equation}
M^2=\left(M_{irr}+\frac{Q^2}{4M_{irr}}\right)^2+\frac{J^2}{4M_{irr}^2},
\end{equation}
which pushes the efficiency of the Penrose process up to $0.5$ \cite{Christodoulou:1972kt}. \par
The fact that a Penrose process cannot reduce the irreducible mass of a black hole is a particular consequence of Hawking's area theorem, discussed below. \par
The Penrose process has a corresponding phenomenon in wave scattering on a stationary axisymmetric black hole background known as superradiant scattering \cite{Misner:1972kx,Press:1972zz,Starobinsky:1972am}. Similar effects were already studied in \cite{Zeldovich:1971am,Zeldovich:1971le} where scalar waves incident on a rotating cylinder were examined. For a qualitative understanding of superradiant scattering consider the scalar wave equation $\nabla^{a}\nabla_{a} \Phi =0$ on a Kerr background. It was shown in \cite{Carter:1968rr} by studying the Hamilton--Jacobi equation for a test particle that this equation is separable, therefore $\Phi$ can be written as: $\Phi=\E^{\I(m\phi-\omega t)}R(r)P(\theta)$ where $P(\theta)$ is a spheroidal harmonic. The solutions for $R(r)$ were studied in detail in \cite{Teukolsky:1973ha}. Suitable boundary conditions for $\Phi$ read
\begin{equation}
\Phi(r)=\begin{cases} \E^{-\I(\omega-m\Omega)r_*} \qquad &r\rightarrow r_+ \\ A_{out}(\omega)\E^{\I\omega r_*}+A_{in}(\omega)\E^{-\I\omega r_*} \qquad &r\rightarrow \infty \end{cases}
\end{equation}
where $r_*$ denotes the tortoise coordinate for the Kerr spacetime. The choice of boundary condition at the horizon $r\rightarrow r_+$ is motivated by the requirement that physical observers at the horizon 
should see exclusively ingoing waves. The Wronskian determinant for this solution and its complex conjugate evaluated in both limits leads to
\begin{equation}
|R|^2=1-\left(1-\frac{m\Omega_H}{\omega}\right)|T|^2.
\end{equation}
Therefore, superradiance is observed for $\omega <m\Omega_H$. Interestingly, the amplification of the incoming amplitude depends on the spin of the incident wave \cite{Teukolsky:1974yv,Starobinsky:1973ch}: 0.003 for a scalar wave, 0.044 for an electromagnetic field and 1.38 for gravitational waves. Half-integer fields do not appear, as fermions show no superradiant scattering behavior. This can be understood from the exclusion principle which allows only one particle in each outgoing mode and thus prevents an enhancement of the scattered wave \cite{Unruh:1973ne,Hawking:1974sw}. \par
The occurrence of superradiant scattering in quantum mechanics is well known from the Klein paradox \cite{Klein:1929zz,Dombey:1999id,Wald:1984rg}. The Klein paradox describes the quantum effect that a wave incident on a step potential is reflected with a coefficient $|R|>1$ for a particular relation between potential height and energy of the incident wave. This effect is attributed to pair creation in the strong electric field near the potential step. Therefore, the presence of superradiant scattering in a black hole background suggests the occurrence of particle creation as was already noted in \cite{Starobinsky:1972am,Zeldovich:1971am,Zeldovich:1971le,Teukolsky:1974yv} and later famously shown by Hawking \cite{Hawking:1974rv} (cf. next section). 

\runinhead{The area theorem.}
The above mechanisms of energy extraction are closely related to the important area theorem. The area theorem and the four laws of black hole mechanics rely on a couple of earlier theorems, which are described in the following with no intention of mathematical rigor (cf. the standard reference \cite{Hawking:1973uf} for details).   \par
The rigidity theorem shows under suitable assumptions that the event horizon of a stationary black hole is a Killing horizon. This result can be proven in two independent ways. Carter showed that the horizon of a static black hole is normal to the static Killing vector $\xi^a$, and the horizon of a stationary black hole is normal to the linear combination $\chi^{a}=\xi^{a}+\Omega_H \phi^{a}$ under the assumption of $t-\phi$ orthogonality \cite{Carter:1973lh}. Here $\Omega_H$ denotes the angular velocity of the horizon and $\phi^{a}$ is the Killing vector generating the axial symmetry. In the second proof, Einstein's equations are assumed in order to show that the event horizon of every stationary black hole in vacuum or electrovacuum is a Killing horizon \cite{Hawking:1973uf}.\par 
The Penrose theorem proven in \cite{Penrose:1968ar} states that the null geodesics generating the horizon may have past end points but no end points in the future. In particular, no caustics of the generators can occur when extended into the future. A consequence of this theorem is that black holes cannot bifurcate or vanish \cite{Hawking:1973uf}.\par
The focusing theorem follows from the Raychaudhuri equation for lightlike congruences. It states that, given a positive convergence at any point of the congruence, the cross-section of the beam vanishes in a finite distance provided the weak energy condition and Einstein's equations hold. \par The area theorem follows from the two latter statements: If the lightlike generators of the horizon had a positive convergence at any point, a caustic would occur in finite distance which is forbidden by Penrose's theorem. Therefore, the area of the horizon cannot decrease \cite{Hawking:1971tu,Hawking:1971vc}. The only possibility to evade this conclusion is the presence of a naked singularity, i.e. a singularity not shielded by a horizon. Thus, the presence of such singularities must be excluded by adopting the cosmic censorship conjecture \cite{Penrose:1969pc}. In summary, the prerequisites of the area theorem put strong restrictions on both causal structure (cosmic censorship conjecture) and matter (weak energy condition) in spacetime. In particular, the latter is in general not met when quantum theory is taken into account \cite{Bekenstein:1973mi}.
\par
The area theorem provides an explanation for the bound on the Penrose process (\ref{eq:done1}), since $M_{irr}$ is proportional to the area of the Kerr black hole \cite{Hawking:1971tu,Hawking:1971vc}. Similarly, the need for superradiant scattering of waves as described above and the existence of a spin-spin interaction between a Kerr black hole and a spinning particle can both be seen just from the area law \cite{Bekenstein:1973mi,Wald:1972sz}. The argument for superradiant scattering from the area theorem breaks down for fermions since the respective energy-momentum tensor does not obey the weak energy condition \cite{Unruh:1973ne,Hawking:1974sw}.\par
\runinhead{The four laws of black hole mechanics.}
The work of this decade culminated in the famous four laws of black hole mechanics by Bardeen, Carter and Hawking \cite{Bardeen:1973gs}. These laws, which show a remarkable similarity to the laws of thermodynamics, are stated in the following. 
\begin{itemize}
\item{The zeroth law of black hole mechanics states that the surface gravity $\kappa$ is constant on the horizon of a black hole. The proof of this result as given in \cite{Bardeen:1973gs} requires the dominant energy condition to hold and the use of Einstein's equations. Similar results were already obtained in \cite{Carter:1971zc,Hawking:1971vc}. Another proof was given in \cite{Carter:1973lh} by using the assumption of $t-\phi$ orthogonality \cite{Carter:1973lh}. The zeroth law suggests a similarity between $\kappa$ and the temperature of a body in thermal equilibrium.}
\item{
The first law establishes a relation between changes in the mass $M$, horizon area $A$, angular momentum $J_H$, and charge $Q_H$ if the black hole is perturbed.
\begin{equation}
\delta M =\frac{\kappa}{8\pi}\delta A+ \Omega_H \delta J_{H}+\Phi_H \delta Q \label{eq:done3}
\end{equation}
In \cite{Bardeen:1973gs} the first law was derived from a generalized version of the Smarr mass formula \cite{Smarr:1972kt}. The first law bears a clear resemblance to the first law of thermodynamics with $\kappa$ as temperature, the horizon area $A$ taking the place of entropy, and the mass $M$ taking the role of energy. As pointed out in \cite{Wald:1994qf}, in fact two different versions of the first law exist: an equilibrium version, wherein one compares the parameters of neighboring equilibrium solutions, and a physical process version in which the parameters of the black hole are changed, e.g. by dropping in matter, and analyzing the change in the parameters after the black hole has settled down. These two independent versions yield the same result.   }
\item{The second law of black hole mechanics is Hawking's area theorem:
\begin{equation}
\delta A \ge0.
\end{equation}
Here, the analogy between horizon area $A$ and entropy becomes evident.}
\item{The third law states that the surface gravity of a black hole cannot be reduced to zero in a finite number of processes. This formulation is an analogue of the Nernst unattainability principle \cite{Nernst:1912th}. The Planck formulation of the third law of thermodynamics does not hold in black hole mechanics, as the horizon area of an extremal black hole is finite despite vanishing surface gravity. It follows from the third law that non-extremal black holes cannot be made extremal in a finite number of steps. A proof for the third law was presented later in \cite{Israel:1986tl}. }
\end{itemize}
The close mathematical and physical analogy between the four laws of black hole mechanics and the laws of thermodynamics is remarkable. Nonetheless, it appears to be a mere analogy in classical general relativity. Classical black holes do not have temperature since they cannot radiate, and entropy is a dimensionless quantity in contrast to the horizon area that has a dimension of length squared. It is only when quantum theory is taken into account that the analogy becomes an identity.

\section*{1973--1983}
 
 
\runinhead{Bekenstein--Hawking entropy.}  In 1971, when Wheeler proposed the now famous gedankenexperiment of pouring a hot cup of tea into a black hole, he was questioning whether black holes violate the second law of thermodynamics.
Another possible violation of the second law of thermodynamics by classical black holes was put forward by Geroch: A box of matter with mass $m$ is lowered close to the horizon of a black hole from infinity where its energy (as measured from infinity) is nearly zero, thus providing an amount $m$ of work. At the horizon the box radiates away an amount $\delta m$ of its mass and is hauled back to infinity, requiring an amount $m-\delta m$ of work. In this process an amount $\delta m$ of heat is transformed entirely into work thereby violating the second law of thermodynamics. \par
These evident violations of the second law of thermodynamics and Hawking's area theorem led Bekenstein to propose an entropy for black holes that is proportional to horizon area measured in units of Planck area with a coefficient $\eta$ of order one \cite{Bekenstein:1972tm,Bekenstein:1973ur}:
\begin{equation}
S_{BH}=\eta \frac{k_Bc^3 A}{\hbar G}. \label{eq:dtwo1}
\end{equation} 
The second law of thermodynamics is then replaced by a generalized second law of thermodynamics which states that the change in the sum of matter entropy and black hole entropy is strictly nonnegative. Bekenstein showed that this generalized second law resolves the problems associated with the Geroch process \cite{Bekenstein:1973ur}. Furthermore, he tested the law for the cases of a harmonic oscillator enclosed in a spherical box and infalling radiation.\par

\runinhead{Hawking radiation.}
The relation between black hole entropy and horizon area together with the first law of black hole mechanics indicates that black holes do have a temperature that should be proportional to $\kappa$. 
If the black hole is immersed in black body radiation of lower temperature then the generalized second law is violated, unless the black hole also emits radiation. Therefore, spontaneous particle creation is needed to prevent a violation of the generalized second law. Eventually, Hawking showed that black holes spontaneously emit radiation characteristic of a black body at temperature 
\begin{equation}
T_{H}=\frac{\hbar \kappa}{2\pi c k_B},
\end{equation}
thus establishing also further evidence for the validity of the generalized second law \cite{Hawking:1974sw,Hawking:1974rv}. The coefficient $\eta$ in (\ref{eq:dtwo1}) is fixed to $\eta=\frac{1}{4}$ as can be seen from the first law of black hole mechanics (\ref{eq:done3}). Consequently, black holes can be treated as thermodynamic systems, and the four laws of black hole mechanics cease to be mere analogies, but describe black holes as thermodynamic systems. \par

In the original derivation of Hawking radiation a massless scalar field is studied in the background metric of gravitational collapse. The scalar field yields a decomposition in terms of a complete set of solutions both on lightlike past infinite, $\mathcal{I}^{-}$, and on the union of the horizon and lightlike future infinity, $\mathcal{I}^{+}$. 
Both sets of solutions contain positive frequency modes with respect to the appropriate affine parameters on $\mathcal{I}^{+}$ and $\mathcal{I}^{-}$. The different decompositions of the field induce a Bogoliubov transformation on the two sets of creation and annihilation operators on $\mathcal{I}^{+}$ and $\mathcal{I}^{-}$.
Therefore, the vacuum state with respect to the operators for the ingoing particles yields a nonzero expectation value for the number operator for an observer in $\mathcal{I}^{+}$.\footnote{
In quantum information language the vacuum quantum state in a black hole space-time for each mode is
a two-mode squeezed vacuum, similar to what happens for primordial density fluctuations in cosmology \cite{Grishchuk:1990bj}.} This is the Hawking radiation. The particle number measured at $\mathcal{I}^{+}$ of a particular mode is given by \cite{Hawking:1974sw}
\begin{equation}
\langle n_J \rangle=\frac{\Gamma_J}{\exp\left[(2\pi \bar{\omega}_J)/\kappa\right]\pm1}, \label{eq:dtwo2}
\end{equation}
where the index $J$ denotes collectively frequency $\omega$, angular momentum $l$, azimuthal quantum number $m$, sign of the charge, and spin. The upper sign is for fermions and the lower sign for bosons. Here the quantity $\bar{\omega}$ is defined as $\bar{\omega_J}=\omega_J-m_j\Omega_H-q \Phi_H$, and $\Gamma_J$ denotes the fraction of the incident radiation that enters the collapsing body, i.e. $\Gamma_J=1-|R_J|^2$. Expression (\ref{eq:dtwo2}) is precisely the result expected for a black body with temperature $\kappa/(2\pi)$ and greybody factor $\Gamma_J$. In fact, Hawking radiation is completely identical to black body radiation, since the density matrices for Hawking radiation and black body radiation coincide \cite{Wald:1975kc,Parker:1975jm,Hawking:1976ra}. Furthermore, it was shown that black holes behave like black bodies even in the presence of incoming radiation. Expressions for the probability of emission of $k$ particles when $m$ particles have arrived, $P(k|m)$, and the Einstein coefficients for induced emission, spontaneous emission and absorptions were obtained in \cite{Bekenstein:1977mv,Panangaden:1977pc}. The derivation of Hawking was repeated subsequently in various approaches and generalizations (cf.\cite{Boulware:1974dm,Gibbons:1978kq,Davies:1974th,DeWitt:1975ys,Gerlach:1976ji}).\par
The Hawking effect is often described heuristically as Schwinger pair creation in the gravitational field of a black hole, where the negative energy particle drops into the black hole and the other particle escapes to infinity \cite{Hawking:1974sw}. A derivation of the Hawking effect that closely resembles this picture of a tunneling process was presented in \cite{Parikh:1999mf}.

\runinhead{Hawking radiation from anomalies.}
Particularly striking is the connection between Hawking radiation and anomalies of the stress-energy tensor. If restricted to the s-wave sector, Hawking radiation can be studied in an effectively two dimensional spacetime. In this geometry, Hawking radiation can be shown to arise from the trace anomaly of the energy-momentum tensor for a massless field, by requiring finiteness of $T_{\mu \nu}$ at the horizon for a geodesic observer \cite{Christensen:1977jc} (cf. e.g. \cite{Strominger:1994lh} for a general discussion, and \cite{Grumiller:2002nm}).\par
More recently, it was shown that Hawking radiation is necessary for the cancellation of gravitational anomalies \cite{Robinson:2005pd}---i.e. non-conservation of $T_{\mu \nu}$---in Schwarz\-schild spacetimes of any dimension. A gravitational anomaly occurs if one assumes that modes propagating along the horizon can be integrated out, so that $T_{\mu\nu}$ is regular on the horizon. Thus, the resulting theory is effectively chiral near the horizon and acquires a gravitational anomaly, which is removed by Hawking radiation. This method can be generalized to charged and rotating black holes \cite{Iso:2006ut,Iso:2006wa}.\par
\runinhead{Euclidean path integral.}
Due to its intimate connection with the partition function, the Euclidean path integral formalism of quantum gravity is widely used when studying black hole thermodynamics. \par
The singularities encountered in black hole spacetimes can be avoided by a Wick rotation into the Euclidean sector. This requires a periodic Euclidean time with periodicity of inverse temperature. In general, the Euclidean path integral does not converge due to the presence of the conformal mode \cite{Gibbons:1978ac}. However, in the semi-classical approximation to the partition function, i.e. expanding the path integral around solutions of the classical  equations of motion, the above results for entropy and temperature of black holes are recovered \cite{Gibbons:1976ue,Hawking:1978jz}.\par
In the case of flat spacetime at non-zero temperature studied in \cite{Gross:1982cv}, a sum over the Schwarzschild instanton in the path integral leads to a non-zero probability for the decay of flat space into a black hole. Gravitational instantons and their thermodynamic properties were studied and classified in \cite{Gibbons:1979xm,Gibbons:1979nf,Gibbons:1978ji}.\par

\runinhead{Modes of black hole decay.}
It is seen from (\ref{eq:dtwo2}) that the emission of particles with charge of the same sign as the charge of the black hole is enhanced. Thus, the charge of the black hole is radiated away \cite{Zaumen:1974an,Carter:1974yx,Gibbons:1975kk,Page:1976df,Page:1977um}. The resulting current is proportional to the particle number (\ref{eq:dtwo2}) times the charge of the emitted particle. An estimation of the discharge rate yields that the timescale over which the discharge occurs is in general much shorter than the relevant timescale for formation of the black hole \cite{Carter:1974yx}, provided that $Q/M \ge M m_e^2/e$. Thus, only very large black holes show a significant charge $Q$. All other black holes show only random charge fluctuations of order $(\hbar c)^{1/2}$ after sufficiently long time \cite{Page:1976df,Page:1977um}.\par
For fixed angular momentum $l$ the emission of particles with positive azimuthal quantum number $m$ is enhanced, and the black hole loses angular momentum. When radiation of massless particles only is considered, the black hole loses angular momentum considerably faster than mass \cite{Page:1976ki}. Curiously, the emission of neutrinos shows parity violation: antineutrinos are emitted preferentially parallel to angular momentum whereas more neutrinos are emitted in the opposite direction \cite{Unruh:1974bw,Vilenkin:1979ui,Leahy:1979xi}.\par
Radiation of the black hole mass occurs over a timescale $\tau \propto G^2 M_0^3/(\hbar c^{4})$, which exceeds the age of the present universe unless the black hole is sufficiently light, $M_0\le 5 \times 10^{11} kg$. The species of the emitted particles changes with the mass of the black hole: black holes emit massless particles only as long as $M\ge 10^{14}kg$ at which point electron-positron emission starts; the onset for emission of heavier particles lies at $M\approx 10^{11} kg$ \cite{Page:1976df,Page:1976ki,Page:1977um}. Consequences of black hole evaporation for unitarity are discussed below. \par

\runinhead{Unruh effect.}
The Unruh effect describes the detection of vacuum fluctuations of the Minkowski vacuum as thermal radiation by a constantly accelerated observer, i.e. an observer in Rindler spacetime \cite{Davies:1974th,Fulling:1972md,Unruh:1976db,Unruh:1983ac}. The Minkowski vacuum--- the vacuum for an observer measuring time along the Killing vector $\partial_t$---can be represented as the sum
\begin{equation}
|0\rangle=\sum_{n}\exp{(-2\pi a^{-1} \omega_n)}|n\rangle_L\times |n\rangle_R.
\end{equation}
where $|n\rangle_L (|n\rangle_R)$ are states with energy $\omega_n$ measured by an observer moving with acceleration $a$ along the respective Killing vectors in the left (right) wedge of Rindler spacetime \cite{Unruh:1983ac,Fulling:1987tp}.\footnote{The sum should be regarded as formal since the quantum theory constructions of the two observers are unitarily inequivalent \cite{Wald:1994qf}.} The Minkowski vacuum thus contains correlations between states in different wedges of Rindler spacetime and is regarded as a thermal bath of temperature
\begin{equation}
T_{\textrm{U}}=\frac{\hbar a}{2\pi c k_B}
\end{equation}
by the accelerating observer. Although the original derivation was given for free fields, the validity of the Unruh effect for interacting fields is a consequence of general results obtained in axiomatic quantum field theory \cite{Bisognano:1975ih,Bisognano:1976za}, as first recognized in \cite{Sewell:1982zz}. The Unruh effect indicates that already in flat spacetime the notion of particles is observer dependent. Other seemingly paradoxical aspects of the Unruh effect are covered in \cite{Unruh:1983ms}. Recent developments and issues regarding experimental detection are reviewed in \cite{Crispino:2007eb}.\par
The Unruh effect is also invoked to prevent a violation of the generalized second law in the following form: A box with given energy and entropy is released from infinity and its content dropped into the black hole. The energy gain of the black hole can be made arbitrarily small by dropping the box from a point close to the horizon. The horizon area might not increase enough to compensate the loss of entropy, thus violating the generalized second law. In \cite{Bekenstein:1980jp}, a universal upper bound on the ratio entropy to energy was proposed, which would prevent such violations of the generalized second law. On the other hand, it was argued that the box would feel an effective buoyancy force near the black hole originating from the acceleration radiation. This buoyancy force guarantees a lower bound on the energy gain of the black hole, thus saving the generalized second law without the need for an entropy bound \cite{Unruh:1982ic,Unruh:1983ir}. 
\par
The similarities between Hawking and Unruh effect are due to the similar horizon structure: any non-extremal Killing horizon looks like a Rindler horizon in the near-horizon approximation. 
Depending on the choice of boundary conditions different vacua exist, which are suitable for different physical applications. 
The Unruh vacuum fixes boundary conditions on the past horizon $H^{-}$ and $\mathcal{I}^{-}$ \cite{Unruh:1976db}. This state is analogous to the original treatment of black hole evaporation by Hawking. For the Hartle--Hawking vacuum one defines boundary conditions on both future and past horizon $H^{+}$ and $H^{-}$ \cite{Hartle:1976tp,Israel:1976ur}, which describes a black hole in equilibrium with incoming radiation, and is therefore the relevant state for the curved spacetime generalization of the Unruh effect. This state does not exist for Kerr black holes \cite{Kay:1988mu}. The Boulware vacuum sets boundary condition on $\mathcal{I}^{+}$ and $\mathcal{I}^{-}$ and describes a state with no radiation \cite{Boulware:1974dm}, but is singular on past and future horizon and therefore of little physical significance. \par

\runinhead{The transplanckian problem and black hole analogue systems.}
Since the direct experimental verification of black hole thermodynamics effects is (and most likely will remain) out of reach, analog systems have been proposed in which the Hawking effect could be studied. One of the first proposed systems concerns sound waves in a convergent fluid flow \cite{Unruh:1981ee}. The linearized equations of motion correspond to the equations for a massless scalar field in a background metric that can be brought in a Schwarzschild-like form, thus producing a sonic black hole with the speed of light replaced by the speed of sound. Quantization of the scalar field in this background leads to the emission of sound waves in a thermal spectrum at the sonic horizon, the temperature of which is given by a quantity analogous to the Hawking temperature. Albeit very small, this quantity should be measurable in principle. The field of analogue gravity has grown rapidly in the last decades; the interested reader is referred to \cite{Novello:2002qg,Barcelo:2005fc} and references therein. \par
Black hole analogue systems play an important role in the study of the transplanckian puzzle. A Hawking mode of frequency $\omega$ measured at infinity that was emitted a time $t$ after formation of the black hole stems from a fluctuation of frequency $\omega \exp{(\kappa t)}$. This means that modes emitted a sufficiently long time after the formation of the black hole originated from modes beyond the Planck energy, where the theory can no longer be trusted. This raises the question if the Hawking effect depends on the details of a transplanckian theory. Certainly, Lorentz invariance would guarantee the validity of the derivation, but it is a logical possibility (though one that is highly-constrained by observations from the Fermi Large Area Telescope \cite{Ackermann:2009aa}) that Lorentz invariance is broken at arbitrarily high energies, see \cite{Jacobson:1991gr} for a discussion. A viable option, at least for analogue systems, is the study of Hawking radiation with a modified dispersion relation at high frequencies. Since this situation is similar to the study of black hole analog systems in fluid mechanics, where the theory breaks down at wavelengths comparable to the atomic scale, these systems are used in the study of the transplanckian problem. A particular example was presented in \cite{Unruh:1994je}, where it was shown that Hawking radiation occurs despite a change in the dispersion relation at high frequencies. \par
The transplanckian problem was far from being settled in that decade, but it seems that Hawking radiation is robust enough to persist, even if the theory is modified at ultrahigh energies like in analogue systems \cite{Corley:1996ar,Jacobson:1993hn,Brout:1995wp,Visser:1997yu,Jacobson:1996zs,Visser:1997ux}. \par

\runinhead{Black hole evaporation and information loss.}
As pointed out above, black holes evaporate due to Hawking radiation on a timescale $\tau \propto G^2 M_0^3/(\hbar c^{4})$, which is of order $10^{70}s$ for a solar mass black hole. Although this amount of time is enormous already for solar mass black holes, the very fact that black holes evaporate reveals the deep conceptual problem of information loss, first raised in \cite{Hawking:1976ra}. At the classical level, the no-hair theorem implies that the large amount of data needed to describe the precollapse geometry is reduced to a small number of quantities that describe the black hole. The remaining information of the precollapse geometry is not accessible to the outside observer, but in principle can be thought of as residing in the black hole. The real paradox rears its head when Hawking radiation is taken into account. Consider an initial pure state that describes an object falling into the black hole. The Hawking radiation emitted by the black hole is in a mixed state due to correlations between states outside the horizon and states inside the black hole, but after some time the black hole has evaporated completely, and one is left only with the mixed state of Hawking radiation. The evolution from the initial pure state to perfectly thermal Hawking radiation is therefore not unitary and information appears to be lost in the process. This is in contrast to ordinary physical systems like a star or a burning lump of coal, where the emissions contain correlations that would in principle allow one to reconstruct the initial state. It was not clear at the time if this might also be a viable explanation for an evaporating black hole, mostly due to the lack of a sufficiently detailed theory of quantum gravity.

\section*{1983--1993}


The results of the previous decade revealed several problems -- the information paradox, the universality of the area law, and the nature of the states underlying the Bekenstein-Hawking entropy -- that became the focus of research during the period 1983 - 1993. Many researchers turned their attention towards lower-dimensional models, where theories are more tractable but still suffer from conceptual issues such as the information paradox. At the same time, investigations into a diverse array of gravitational theories revealed certain universal features of black hole thermodynamics and led to the first early successes in a state-counting approach to explaining black hole entropy.
Before delving into lower-dimensional gravity we state some of the main conclusions that were reached from its study.

\runinhead{What to do with information loss?}
The information loss problem is of conceptual rather than technical nature. 
Like other conceptual issues in classical and quantum gravity, it arises independently from the spacetime dimension.
Therefore, a useful strategy is to consider lower-dimensional models of gravity where the technical problems become more manageable, conceptual issues can be addressed and, ideally, resolved. See \cite{Brown:1988} for a textbook on lower-dimensional gravity from 1988. 
Particularly the CGHS model of string-inspired 2-dimensional dilaton gravity with matter \cite{Callan:1992rs} (see below) inspired numerous investigations of evaporating black holes in two dimensions, such as the one by Russo--Susskind--Thorlacius \cite{Russo:1992ax}. Exact solubility (even in the presence of quantum effects) is a key feature of the RST model, which allows to address the endpoint of Hawking evaporation. Depending on the energy flux of the infalling matter either no horizon forms or an apparent horizon does form and eventually evaporates to a naked singularity, which requires the imposition of suitable boundary conditions, for which a natural choice exists in this model. Most importantly, the whole process is described in a unitary way, so that all information is recovered in this case.

\runinhead{Black hole complementarity.}
Based on studies of 2-dimensional dilaton gravity models, Susskind, Thorlacius and Uglum advocated the ``black hole complementarity'' principle \cite{Susskind:1993if} (which was formulated independently in \cite{Stephens:1994an}). The essence of this principle is captured by four postulates (three of which were spelled out explicitly in \cite{Susskind:1993if}, which we quote verbatim): 1.~The process of formation and evaporation of a black hole, as viewed by a distant observer, can be described entirely within the context of standard quantum theory. In particular, there exists a unitary S-matrix which describes the evolution from infalling matter to outgoing Hawking-like radiation. 2.~Outside the stretched horizon\footnote{%
The stretched horizon (or the earlier ``brick wall'' \cite{'tHooft:1984re}) is also discussed in these papers and captures the membrane description of a black hole suitable for a distant observer.
} of a massive black hole, physics can be described to good approximation by a set of semi-classical field equations. 3.~To a distant observer, a black hole appears to be a quantum system with discrete energy levels. The dimension of the subspace of states describing a black hole of mass $M$ is the exponential of the Bekenstein entropy \eqref{eq:td1}. 4.~A freely falling observer experiences nothing extraordinary when entering the black hole. 

The attribute ``complementarity'' refers to the fact that the outside observer detects a membrane-like structure near the black hole horizon where information is stored, while the infalling observer sees no membrane at the horizon. The reason why these mutually exclusive viewpoints do not necessarily generate a contradiction is because there should not exist any ``super-observer'' that simultaneously has access to both viewpoints. 

\runinhead{Lower-dimensional gravity.} The lowest spacetime dimension that makes sense to consider is 1+1, since this is the lowest dimension where the notions of black holes, causal structure and curvature exist.
If additionally the existence of graviton excitations (at least off-shell) is required then the lowest dimension one can consider is 2+1, since this is the lowest dimension where linearized perturbations of the metric $h_{\mu\nu}$ have a transverse-traceless part, $h_{\mu\nu}=h_{\mu\nu}^{\textrm{\tiny TT}} + \nabla_{(\mu}\xi_{\nu)} + \tfrac13\,h\,g_{\mu\nu}$. Moreover, 2+1 is the lowest dimension where the notion of the area of the event horizon is meaningful (in 1+1 dimensions this `area' is just a point).
For these reasons, the main focus in lower-dimensional gravity is on 1+1 and 2+1 dimensional models, depending on the scope of the model.

\runinhead{Dilaton gravity in two dimensions.}
In two dimensions there are various ways to motivate which kind of gravity model one should consider. The theory {\em not} to consider is Einstein gravity, since there are no meaningful Einstein equations in two dimensions (the 2-dimensional Einstein tensor vanishes trivially for any metric). Instead, there are (at least) five different ways to end up with the same class of models, namely 2-dimensional dilaton gravity. Its bulk action
\begin{equation}
 I = \frac{1}{16\pi G}\,\int\extd^2x\sqrt{-g}\,\big(XR - U(X)(\partial X)^2-2V(X)\big) 
\label{eq:td31}
\end{equation}
depends on two free functions, $U(X)$ and $V(X)$, of the dilaton field $X$. We summarize briefly five different ways to end up with an action of type \eqref{eq:td31}.
\begin{enumerate}
 \item {\bf Gravity as gauge theory.} Jackiw \cite{Jackiw:1984} and Teitelboim \cite{Teitelboim:1984} considered a 2-dimensional gravity model with constant curvature, which can be formulated as a non-abelian BF-theory with gauge group $SO(2,1)$ \cite{Isler:1989hq,Chamseddine:1989yz}. The generators $P_a$ and $J$ are interpreted as translation and boost generators, respectively. They obey the algebra $[P_a,\,P_b]=\Lambda\epsilon_{ab}\,J$ and $[P_a,\,J]=\epsilon_a{}^b P_b$, where $\Lambda$ is a parameter that sets the scale of curvature (one could call it `cosmological constant'). The $so(2,1)$ connection $A=e^a P_a+\omega J$ decomposes into zweibein $e^a$ and (dualized) connection $\omega = \tfrac12\, \omega^{ab}\epsilon_{ab}$. Its non-abelian field strength $F$ is then coupled linearly to co-adjoint scalars in the BF-action, which reads explicitly
\begin{equation}
 I \sim \int \big(X_a (\extd e^a + \epsilon^a{}_b\omega e^b) + X\extd\omega + \epsilon_{ab}\,e^a\wedge e^b\,\Lambda X\big)\,.
 \label{eq:td32}
\end{equation}
 Integrating out the auxiliary field $X_a$ establishes the constraint of vanishing torsion, which allows to eliminate also the spin-connection $\omega$ and to convert the first order action \eqref{eq:td32} into the second order action \eqref{eq:td31} with $U(X)=0$ and $V(X)=\Lambda X$. A similar BF-type of construction was provided by Cangemi and Jackiw \cite{Cangemi:1992bj} for a string inspired model discussed below. The gauge theoretic formulation for arbitrary dilaton gravity theories was provided by Ikeda and Izawa \cite{Ikeda:1993aj,Ikeda:1993fh} and by Schaller and Strobl \cite{Schaller:1994es}, dubbed ``Poisson-$\sigma$ model''. 
 \item {\bf Dimensional reduction.} Assuming spherical symmetry in $D$ spacetime dimensions leads to a line-element in adapted coordinates that depends on a 2-dimen\-sio\-nal metric and a scalar field, $\extd s^2=g_{\alpha\beta}\,\extd x^\alpha\extd x^\beta+X^{1/(D-2)}\,\extd \Omega^2_{S^{D-2}}$, where $\extd \Omega^2_{S^{D-2}}$ denotes the line-element of the round $(D-2)$-sphere \cite{Berger:1972pg, Benguria:1977in, Thomi:1984na, Hajicek:1984mz, Kuchar:1994zk}. Inserting this ansatz into the $D$-dimensional Einstein--Hilbert action permits to integrate out all angular coordinates and eventually establishes an effective 2-dimensional model whose bulk action is precisely \eqref{eq:td31}, with $U(X)=-(D-3)/[(D-2)X]$ and $V(X)\propto X^{(D-4)/(D-2)}$. Curiously, in the limit $D\to\infty$ the model derived from bosonic string theory is recovered (with 2-dimensional target space, see below) \cite{Soda:1993xc,Grumiller:2002nm,Emparan:2013xia}.
 \item {\bf Limiting case of Einstein--Hilbert in $2+\varepsilon$ dimensions.} Weinberg's idea of asymptotic safety in gravity emerged from his consideration of gravity in $2+\varepsilon$ dimensions, in the limit of small $\varepsilon$ \cite{Weinberg:1979}. As we mentioned above, taking $\varepsilon\to 0$ leads to trivial equations of motion. However, if simultaneously Newton's constant scales to zero appropriately, then the limiting action can be non-trivial. In fact, Mann and Ross argued that the action obtained in this way is a 2-dimensional dilaton gravity action \eqref{eq:td31} with $U(X)=\rm const.$ and $V(X)=0$ \cite{Mann:1992ar}. A more recent analysis confirms the result for $U(X)$, but finds $V(X)\propto e^{-2X}$ \cite{Grumiller:2007wb}.\footnote{The derivation in \cite{Grumiller:2007wb} exploits a spherically symmetric ansatz in $2+\varepsilon$ dimensions, dualizes to a different action for which the limit $\varepsilon\to 0$ is well-defined and dualizes back after taking the limit.} Such an action describes Liouville gravity, see \cite{Ginsparg:1993is,Nakayama:2004vk} for reviews.  
 \item {\bf Higher power curvature theories.} Models that are non-linear in curvature and/or torsion are viable in two dimensions. In particular, the Katanaev--Volovich model describes 2-dimensional Poincar\'e gauge theory, i.e., a model with Lagrange density $R^2+T^2$, where $R$ is curvature and $T$ torsion \cite{Katanaev:1986wk}. The Katanaev--Volovich model is classically equivalent to dilaton gravity \eqref{eq:td32} with $U(X)=\rm const.$ and $V(X)\propto X^2$, see \cite{Kummer:1992bg,Schaller:1992np,Ikeda:1993fh}. Similarly, generic theories with non-linear Lagrangians in curvature and torsion are equivalent to generic dilaton gravity, provided the potentials $U(X)$ and $V(X)$ are chosen appropriately \cite{Katanaev:1995bh}. 
 \item {\bf Strings in two dimensions.} Conformal invariance of the sigma model action for the closed bosonic string,
\begin{equation}
 I^{(\sigma)} = \frac{1}{4\pi\alpha^\prime}\,\int\extd^2z \sqrt{-h}\,\big(g_{\mu\nu} h^{ij} \partial_i X^\mu \partial_j X^\nu+\alpha^\prime\,\Phi\,{\cal R}\big)
\label{eq:td33}
\end{equation}
requires that the trace of the world-sheet energy-momentum tensor vanishes
\begin{equation}
 T^i_i \propto \beta^\Phi {\cal R} + \beta^g_{\mu\nu} h^{ij} \partial_i X^\mu \partial_j X^\nu = 0\,.
\label{eq:td34}
\end{equation}
The parameter $\alpha^\prime$ is the string tension, $h_{ij}$ is the world-sheet metric, ${\cal R}$ its Ricci scalar, $X^\mu$ are the target space coordinates, and $\Phi$ is the dilaton field.
Thus, for consistency the $\beta$-functions appearing in \eqref{eq:td34} have to vanish \cite{Callan:1985ia}.
\begin{align}
 \beta^\Phi &= - \frac{\alpha^\prime}{4\pi^2}\,\big(\tfrac{26-D}{12\alpha^\prime} + (\partial\Phi)^2-4\nabla^\mu\partial_\mu\Phi-\tfrac14\,R\big) = 0\\
 \beta^g_{\mu\nu} &= R_{\mu\nu} + 2 \nabla_\mu\partial_\nu\Phi = 0
\end{align}
Here $D$ is the dimension of the target space, $R_{\mu\nu}$ its Ricci tensor and $\nabla_\mu$ the associated covariant derivative.
The conditions of conformal invariance, $\beta^\Phi=\beta^g_{\mu\nu}=0$, follow as equations of motion from a target space action, which for $D=2$ turns out to be equivalent to dilaton gravity \eqref{eq:td31} with $U(X)=-1/X$ and $V(X)=2\lambda^2 X$, upon identifying $X=e^{-2\Phi}$. See \cite{Mandal:1991tz,Elitzur:1991cb,Witten:1991yr,Dijkgraaf:1992ba} for some early literature on black holes in 2-dimensional string theory and \cite{Klebanov:1991qa,Ginsparg:1993is} for some reviews. The model by Callan, Giddings, Harvey and Strominger (CGHS) uses the same target space action as derived from string theory and adds matter fields to describe evaporating black holes \cite{Callan:1992rs}; the CGHS model engendered a lot of further research in 2-dimensional dilaton gravity with and without matter, see \cite{Strominger:1994tn,Grumiller:2002nm,Grumiller:2006rc} for reviews. 
\end{enumerate}
Thermodynamics of 2-dimensional dilaton gravity models \eqref{eq:td31} was discussed assuming $U(X)=0$ by Gegenberg, Kunstatter and Louis-Martinez \cite{Gegenberg:1994pv}. A comprehensive discussion of quasi-local thermodynamics for generic models \eqref{eq:td31} was provided using the Euclidean path integral approach more than a decade later \cite{Grumiller:2007ju}. 

Taken together, the body of results that these diverse two-dimensional models have in common suggests that certain features of black hole thermodynamics are universal. This is an important observation in its own right, independent of insights into the information paradox and other problems. In particular, with appropriate normalizations the `classical' contribution to the entropy always takes the form
\begin{align}\label{eq:2DEntropy}
	S = 2\pi X_{h}
\end{align}
where $X_h$ is the value of the dilaton at the horizon. This result encapsulates inherently two-dimensional models, as well as the s-wave reduction down to two dimensions of the area law for higher dimensional theories. The robust nature of black hole entropy was made apparent in the work of Wald, who gave a succinct geometric characterization of the entropy for any diffeomorphism-invariant theory of gravity \cite{Wald:1993nt,Iyer:1994ys}.

\runinhead{Quasi-local thermodynamics and Hawking--Page phase transition.} A simple calculation shows that the Schwarzschild black hole has a negative specific heat, and therefore cannot be treated as an equilibrium thermodynamic system.  
York addressed the issue of negative specific heat by putting the black hole inside a cavity of some finite radius that provides a heat bath of fixed temperature \cite{York:1986it}. For a sufficiently small cavity the specific heat is positive, leading to a well-defined canonical ensemble. It was shown later that some spacetimes, in particular asymptotically AdS spacetimes, naturally provide a covariant version of such a cavity. In all these examples the existence of a well-defined canonical ensemble means that interesting phase structures can be unraveled. Probably the most famous example is the Hawking-Page phase transition between ``hot AdS'' --- anti-de Sitter space with periodic euclidean time $\tau \sim \tau + T^{-1}$ --- and an asymptotically AdS black hole \cite{Hawking:1982dh}. For sufficiently small temperatures the minimum of the free energy is hot AdS, while at high temperatures the ensemble is dominated by the black hole.

\runinhead{Gravity in three dimensions and a connection with conformal field theory.}
During the same period there was a great deal of pioneering work in 3-dimensional gravity. 
Deser, Jackiw and Templeton constructed topologically massive gauge theories by adding a Chern--Simons term to the action \cite{Deser:1982vy,Deser:1982wh,Deser:1982a}. In the case of gravity this leads to topologically massive gravity, a 3-dimensional theory of gravity that has a local (massive) gravitational degree of freedom. Its bulk action reads
\begin{equation}
16\pi G\,I^{\textrm{\tiny TMG}} = \int\extd^3x\sqrt{-g}\,\big(R-2\Lambda\big) +  \frac{1}{2\mu}\,\int\extd^3x \,\varepsilon^{\mu\nu\lambda}\,\Gamma^\alpha{}_{\mu\beta}\,\big(\partial_\nu\Gamma^\beta{}_{\lambda\alpha} + \tfrac23\,\Gamma^\beta{}_{\nu\gamma}\Gamma^\gamma{}_{\lambda\alpha}\big)
\label{eq:td35}
\end{equation}
Without the gravitational Chern--Simons term, $\mu\to\infty$, Einstein gravity becomes locally trivial \cite{Deser:1983tn}, but globally it can be non-trivial. 
In particular, in a seminal paper Brown and Henneaux found that the Hilbert space of any 3-dimensional theory of quantum gravity with AdS boundary conditions falls into representations of two copies of the Virasoro algebra, with central charges for Einstein gravity given by \cite{Brown:1986nw}
\begin{equation}
c = \bar c = \frac{3\ell}{2G}\qquad \textrm{where}\; \Lambda = - \frac{1}{\ell^2}\,.
 \label{eq:td36}
\end{equation}
This unexpected set of symmetries suggested that such theories might be amenable to an analysis using conformal field theory (CFT) techniques. The Brown--Henneaux results were an important precursor of the AdS/CFT correspondence found a decade later. 

\runinhead{Black holes in three dimensions.}
Another crucial development was the discovery, by Ba\~nados, Teitelboim and Zanelli (BTZ), of black hole solutions of 3-dimensional Einstein gravity with negative cosmological constant \cite{Banados:1992wn}. As discussed in \cite{Banados:1992gq}, the BTZ black holes are locally AdS, but globally differ from AdS. In fact, they are certain orbifolds of AdS such that the ensuing solutions are locally AdS and remain regular on and outside the event horizon. The line-element in `Boyer--Lindquist' type coordinates ($\varphi\sim\varphi+2\pi$),
\begin{equation}
\extd s^2_{\textrm{\tiny BTZ}} = -\frac{(r^2- r_+^2)(r^2 - r_-^2)}{r^2 \ell^2}\, \extd t^2 +  \frac{r^2 \ell^2}{(r^2- r_+^2)(r^2 - r_-^2)}\, \extd r^2 + r^2\, \big(\extd \varphi + \frac{r_+ r_-}{\ell r^2} \extd t\big)^2
 \label{eq:td37}
\end{equation}
makes the similarity to rotating black holes in higher dimensions manifest: there is an ergosphere at $r=(r_+^2+r_-^2)^{1/2}$, an outer horizon (with rotation) at $r=r_+$, an inner horizon at $r=r_-$, and a singularity behind the inner horizon. Moreover, there is a conserved mass, $M=(r_+^2+r_-^2)/(8G\ell^2)$, and angular momentum, $J=r_+r_-/(4G\ell)$. The presence of rotating black holes makes 3-dimensional AdS gravity a particularly interesting toy model to address classical and quantum aspects of black holes and their thermodynamical properties. In particular, the entropy is given by the Bekenstein--Hawking result \eqref{eq:td1}
\begin{equation}
S_{\textrm{\tiny BTZ}} = \frac{2\pi r_+}{4G} ~.
 \label{eq:td38}
\end{equation}

\runinhead{Cardy formula.}
The existence of the BTZ solution and the results of Brown and Henneaux led to the first attempt to explain black hole entropy by counting microscopic states. Since the Hilbert space of the theory is organized according to the symmetries of a two-dimensional CFT, one can carry out the state counting by exploiting a result of Cardy \cite{Cardy:1986ie,Bloete:1986qm}. Namely, given some assumptions there is a universal formula for the asymptotic density of states in a CFT$_2$. The log of the density of states leads to the Cardy formula for entropy
\begin{equation}
S_{\textrm{\tiny Cardy}} = 2\pi\sqrt{\frac{c h}{6}} + 2\pi\sqrt{\frac{\bar c \bar h}{6}} ~,
 \label{eq:td39}
\end{equation}
where $c$, $\bar c$ are the central charges and $h,\bar h$ are the Virasoro zero-mode charges. Evaluating the Cardy formula \eqref{eq:td39} for the Brown--Henneaux central charges \eqref{eq:td36} and the zero-mode charges $h=(\ell M+J)/2$, $\bar h=(\ell M-J)/2$ of the BTZ black hole \eqref{eq:td37} gives precisely the Bekenstein--Hawking entropy \eqref{eq:td38}. This observation was the basis for the near horizon microstate counting pioneered by Strominger and Vafa a decade later \cite{Strominger:1997eq,Carlip:1998wz}.

The explanation of \eqref{eq:td38} via microscopic state counting was a significant insight into the nature of black hole entropy. But it also left many important questions unanswered. In particular, the Cardy formula provides information about the asymptotic density of states but it gives no insight into the states themselves. An explanation of black hole entropy that proceeds from the \emph{identification} of microscopic states would not be achieved until the following decade.

\runinhead{Towards holography.}
Counting black hole microstates through a CFT calculation is a remarkable manifestation of an idea that began to emerge at the end of the third decade. In an essay dedicated to Abdus Salam, 't~Hooft postulated that there is no information loss, i.e., the evolution describing collapse and quantum evaporation of a black hole should only incorporate processes that are not at odds with unitarity \cite{'tHooft:1993gx}. From this postulate and the observation that the Bekenstein--Hawking entropy \eqref{eq:td1} scales like the area, 't~Hooft then argued that there could be an equivalent description of the system in terms of an ordinary quantum field theory in one dimension lower.\footnote{'t~Hooft also provided as an example a realization of the holographic principle in terms of some cellular automaton model.} A year later Susskind first coined the expression ``holographic principle'' and pointed out that string theory could be a candidate for a theory of quantum gravity realizing the holographic principle. But this part of the story already belongs to the next decade.

\section*{1993--2003}


As described in the previous section, Cardy's formula relates the Bekenstein-Hawking entropy of the BTZ black hole to the central charge of a two-dimensional CFT. This result foreshadows three major developments during the period 1993 - 2003: a complete accounting in string theory of microscopic states responsible for the entropy of certain black holes, the emergence of 't~Hooft and Susskind's holographic principle, and the development of the AdS/CFT correspondence as a fully-fledged example of holography.
 
\runinhead{Counting black hole microstates in string theory.} String theory is a consistent theory of quantum gravity and is therefore a natural framework for investigating the microscopic origin of black hole entropy. As early as 1993, it was suggested that the density of states in perturbative string theory might be sufficient to explain the Bekenstein-Hawking entropy \cite{Susskind:1993ws, Susskind:1994sm, Sen:1995in}. The main development during this period (and arguably one of the most significant accomplishments of string theory in any period) was Strominger and Vafa's calculation of the density of states for certain supersymmetric black holes \cite{Strominger:1996sh}.

String theory contains both bosonic and fermionic degrees of freedom, with the bosonic sector including multiple $p$-form gauge fields under which black holes may be charged. In the case of supersymmetric black holes these charges completely characterize the horizon, which has an area that is independent of moduli like the string coupling or compactification volumes. The simplest such black holes involve either one or two charges, but such configurations possess either singular horizons or horizons with zero area. The Bekenstein-Hawking entropy is relevant for non-singular horizons with macroscopic area, which requires at least three charges.  Strominger and Vafa considered these sorts of black holes in string theory compactified on the five dimensional product spaces $S^1 \times T^4$ and $S^1 \times K3$. Their construction involves $q_1$ D1-branes wrapping the circle, $q_5$ D5-branes wrapping all five compact dimensions, and massless strings stretched between the branes carrying $n$ units of momentum around the $S^1$. At weak coupling this system is described by a supersymmetric field theory on the worldvolume of the branes, and it is possible to enumerate the states with given charges. The resulting density of states is approximately 
\begin{equation}\label{eq:SVDensityOfStates}
	\rho \approx \exp\left(2\pi \sqrt{q_1\,q_5\,n}\right).  
\end{equation} 
As the gravitational (string) coupling is increased the picture changes, and at strong coupling the appropriate description of the system is a black hole. The horizon of this black hole has area
\begin{equation}
 	A_{H} = 8\pi G\,\sqrt{q_1\,q_5\,n} ~,
\end{equation}
where each of the charges must be large to suppress various types of corrections. Although the descriptions at weak and strong coupling are radically different, the state counting is protected by supersymmetry. So even though the density of states \eqref{eq:SVDensityOfStates} was derived at weak coupling, it still applies in the limit where the system is described by the black hole. To leading order the log of the density of states exactly reproduces the area law
\begin{equation}
	S = \log \rho = 2\pi \sqrt{q_1\,q_5\,n} = \frac{A_{H}}{4 G} ~.
\end{equation}
Thus, this result of Strominger and Vafa provides the first derivation of the Be\-ken\-stein--Hawking entropy that identifies and counts a specific set of microscopic states associated with the parameters describing the macroscopic black hole. Similar calculations have been carried out for supersymmetric black holes in four dimensions \cite{Maldacena:1996gb}, near-extremal black holes \cite{Callan:1996dv, Horowitz:1996fn}, and even certain extremal black holes with broken supersymmetry \cite{Emparan:2006it}. Comprehensive reviews can be found in \cite{Skenderis:1999bs, Peet:2000hn}.
   
Despite the success of this program, there is still no explicit construction of the microstates of non-supersymmetric, non-extremal black holes like the Schwarzschild or Kerr solutions (though, in the latter case progress has been made for the extremal solution \cite{Horowitz:2007xq}). It is also important to point out that while the counting of states is protected by supersymmetry, the states in the strong coupling regime bear no resemblance to the states at weak coupling. In this sense, it is not clear what constitutes the ``states of the black hole''. Indeed, given a generic state in the weakly coupled regime it is not clear what happens as the coupling is increased. It is possible (and with hindsight also plausible) that the states in the strongly coupled regime are free of horizons. This idea has motivated a tremendous amount of work -- the so-called \emph{microstate} and \emph{fuzzball} programs -- which will be discussed in the next section.

\runinhead{Holographic principle.} Around the same time that a stringy origin for the black hole density of states was first being considered, 't~Hooft put forth a radical suggestion: that gravitational physics in $3+1$ dimensions must effectively become $2+1$ dimensional at Planckian scales \cite{'tHooft:1993gx}. Susskind, building off his own work on the role of string theory in explaining the Bekenstein--Hawking entropy, explored the consequences of this idea and dubbed it the ``holographic principle'' \cite{Susskind:1994vu}. This principle is often regarded as synonymous with the Bekenstein--Hawking area law for black hole entropy, but it is in fact a much deeper statement about locality, unitarity, and the nature of quantum gravitational physics.

In its earliest form, the holographic principle was interpreted as a bound on the number of degrees of freedom needed to describe physics in a spatial region. Quantum field theory suggests that any such region contains an infinite number of degrees of freedom associated with the infinite number of harmonic oscillator states possible at each of the infinite number of points in the region. Including gravity changes this counting, since exciting too many of these states would provide enough energy to form a black hole. A better estimate would `coarse grain' space on lengths of order the Planck scale and, at the very least, place an upper limit on the energy contained in any Planck volume to avoid creating a black hole. With these restrictions the number of degrees of freedom scales like the volume $V$ of the region. But this must be a gross over-counting, since black holes could still form on larger scales even if the energy bound on each Planck volume was not saturated. And since the largest black hole that `fits' in the region has an entropy given by $A/4$, it must be that the number of accessible degrees of freedom in a region scales like the area bounding the region rather than its volume.

The conclusion described above forces a choice between locality and unitarity. If all the degrees of freedom predicted by local physics were available in a region of volume $V$, then it would not be possible to accommodate all possible states of the system with the dramatically reduced number of states after gravitational collapse. To preserve unitarity, it must be that physics in any region bounded by a surface of area $A$ is described by no more than $A/4$ degrees of freedom, even in the absence of a black hole.

This early form of the holographic principle depends crucially on the idea that the entropy in a spatial region $V$ is limited by the area of the surface $B = \partial V$ bounding the region
\begin{equation}\label{eq:SEB}	
	S[V] \leq \frac{c^{3}}{4\,G\,\hbar}\,A(B) ~.
\end{equation}
But it was soon realized that this spacelike form of the entropy bound can fail \cite{Fischler:1998st, Easther:1999gk},  leading researchers to attempt a reformulation of the bound in terms of light cones. This program culminated with Bousso's Covariant Entropy Conjecture \cite{Bousso:1999xy}, a covariant generalization of the original bound which replaces the spacelike region $V$ with a null hypersurface. Specifically, given some surface $B$ with area $A(B)$ the light sheet $L(B)$ is the null hypersurface generated by following light rays from $B$ until they begin to expand. The entropy on any light sheet of a surface is then bounded according to 
\begin{gather}\label{eq:CEC}
	S[L(B)] \leq \frac{c^{3}}{4\,G\,\hbar}\,A(B) ~.
\end{gather}
A comprehensive review of the Covariant Entropy Conjecture and the holographic principle in general is given in \cite{Bousso:2002ju}. 

Like other entropy bounds, there is no formal derivation of \eqref{eq:CEC}. Rather, it is a conjecture for which there is strong circumstantial evidence and a lack of counterexamples. Since any derivation of this result would require a complete theory of quantum gravity, it is hoped that the holographic principle will instead provide some guidance as to what such a theory might be. It is tempting, given the form of the bounds \eqref{eq:SEB} and \eqref{eq:CEC}, to assume that the physics interior to a region is somehow encoded on its boundary. The holographic principle offers little direct insight as to whether this is the case, or how it might be accomplished\,\footnote{Such an encoding results in an entropy that scales like the area, which suggests a local and non-gravitational description on the boundary.}. Nevertheless, this assumption, combined with calculations inspired by the work of Strominger and Vafa, leads to a fully realized form of the holographic principle in Maldacena's AdS/CFT correspondence.

\runinhead{AdS/CFT correspondence.} The work of Strominger--Vafa showed how the entropy of certain supersymmetric black holes may be understood via a calculation in a field theory on the world volume of a D-brane bound state. The entropy is not the only quantity that can be explained this way. For instance, absorption cross sections calculated using both the gravity and field theory descriptions are found to agree. This observation inspired similar comparisons for a stack of D3-branes in type IIB string theory \cite{Klebanov:1997kc,Gubser:1997yh,Gubser:1997se}. The agreement between the gravity and field theory calculations for the D3-brane system gives the first pieces of evidence for the AdS/CFT correspondence.

Given a stack of $N$ parallel D3-branes, low energy excitations on the worldvolume are described by a four-dimensional $U(N)$ gauge theory with $\mathcal{N}=4$ supersymmetry \cite{Witten:1995im} and a coupling constant related to the string coupling by $g_{YM}^{2} \sim g_s$. For $N$ large and $g_{YM}^{2}\,N \ll 1$ the theory is well-described by perturbation theory with non-planar diagrams suppressed by factors of $1/N$. On the other hand, the near-horizon geometry of the stack of branes looks like a product space of the form $\textrm{AdS}_5 \times \textrm{S}^{5}$, with both factors having a radius of curvature $\ell$ that satisfies
\begin{equation}
	\ell^{\,4} = 4 \pi g_s N (\alpha')^2 ~.
\end{equation}
The description of the system in terms of gravitational physics requires curvatures to be much smaller than the string scale, $\ell \gg \sqrt{\alpha'}$, which implies $g_s\,N \gg 1$. In other words, the gravitational description can be trusted precisely when the worldvolume field theory is strongly coupled. Maldacena conjectured that these descriptions are in fact \emph{the same}; two sides of a strong-weak coupling duality \cite{Maldacena:1997re}. In this picture the conformal symmetries of the field theory are realized by the $SO(4,2)$ isometries of $\text{AdS}_{5}$, while the R-symmetries are encoded in the $SO(6)$ symmetries of the $\text{S}^{5}$.

The strongest form of Maldacena's conjecture asserts that type IIB string theory with $\text{AdS}_{5} \times \text{S}^{5}$ boundary conditions is completely equivalent to four-dimensional Super Yang-Mills for all values of the parameters $g_s$ and $N$. This is the most tantalizing and least tested form of the correspondence. When $N \to \infty$ at fixed $g_{YM}^{2}\,N$ the duality relates classical string theory to Super Yang-Mills with finite coupling, and many consequences of this form of the conjecture have been tested using unexpected integrability properties of the planar sector of SYM \cite{Beisert:2010jr}. The weakest and most thoroughly examined form of the conjecture follows from letting $g_{YM}^{2}\,N \to \infty$. In that case the gravitational side of the duality reduces simply to type IIB supergravity on $\text{AdS}_{5} \times \text{S}^{5}$, which is equivalent to the strong-coupling limit of SYM. All forms of the duality are manifestly holographic, in the sense that the gravitational physics of a $d+1$ dimensional asymptotically AdS spacetime is encoded in a local field theory on the spacetime's $d$ dimensional conformal boundary.

Maldacena's original conjecture, which includes a number of other brane configurations with low-energy descriptions in terms of various supergravities, has been extended, deformed, and modified in various ways. It has primarily been used to extract useful statements about strongly coupled gauge theories. For instance, correlation functions of operators in the gauge theory can be calculated from the string theory partition function, which in the standard (weak) form of the correspondence is dominated by contributions from saddle points of the supergravity action. The on-shell action can be expressed as a functional of `boundary data' $\phi_0$ for the fields $\phi$ that play the role of sources $J$ for operators ${\cal O}$ in the dual field theory
\begin{gather}
	Z_{\textrm{\tiny sugra}}\big[\phi_0 = \phi|_{\partial \textrm{AdS}}\big] = Z_{\textrm{\tiny CFT}}\big[\phi_{0} = J\big] \sim \langle\exp{\big(\int {\cal O} \phi_0\big)}\rangle_{\textrm{\tiny CFT}}~.
\end{gather}
The full impact of AdS/CFT on the study of strongly coupled gauge theories is beyond both the purpose and scope of this review. But the duality does offer several useful insights into black hole thermodynamics, which we will focus on for the rest of this section. 

Not long after the AdS/CFT correspondence was first proposed, Witten showed how the thermodynamics of an AdS black hole can be understood in terms of the (large $N$) thermodynamics of the dual gauge theory \cite{Witten:1998zw}. In particular, the usual Hawking--Page transition from AdS-Schwarzschild to ``hot AdS'' corresponds to a confining/deconfining phase transition in the dual field theory\,\footnote{The dual field theory at finite temperature is defined on $S^{3}\times S^{1}$ and therefore has compact volume. Nevertheless, a phase transition is possible because the theory is considered in the large $N$ limit.}. This can be seen from the free energy of the two bulk configurations, which when expressed in terms of field theory quantities scales as $F\sim O(1)$ and $F \sim O(N^2)$, respectively, in the confined and deconfined phases.

The AdS/CFT correspondence also illuminates calculations of the entropy of the BTZ black hole, raising Brown and Henneaux's result \cite{Brown:1986nw} from an analogy to an actual counting of states in a dual CFT \cite{Strominger:1997eq,Birmingham:1998jt}. This is especially important for a number of black holes that arise in string theory, which typically have near-horizon geometries of the form $\text{BTZ} \times Y$ for some space (or product of spaces) $Y$. The entropy of these black holes can then be explained via a similar state counting without having to work out the full details in the worldvolume theory. For a review, see \cite{Skenderis:1999bs, Peet:2000hn}.

Perhaps the most important consequence of AdS/CFT for black hole thermodynamics is the idea that a gravitational theory, which presumably includes black holes, is equivalent to a theory that is unitary. There are many ways to interpret such a statement in the context of the information paradox. Since the duality applies to the dynamics of both theories it is tempting to `resolve' the paradox by pointing out that any process on the gravity side -- including the formation and eventual evaporation of a black hole -- is encoded in unitary physics on the field theory side. But this is far from a complete argument. In particular, the unitary evaporation of a AdS-Schwarzschild black hole still forces one to either abandon local Hamiltonian evolution (in the bulk) in a setting where it is expected to be a good description, accept the formation of some sort of macroscopic remnant that remains entangled with the Hawking radiation, or else revisit assumptions about the formation of black holes in string theory \cite{Mathur:2009hf}. One possible resolution is that the weakly coupled D-brane states that are counted in, for example, the Strominger--Vafa calculation do not form horizons as the gravitational coupling is increased. Instead, such states possess significant structure on horizon scales, and the traditional black hole is viewed as a coarse-grained description of the actual states. This possibility, which was mentioned earlier, is the basis for the \emph{microstate} and \emph{fuzzball} programs described in the next section.

The AdS/CFT correspondence is, at present, the most fully realized implementation of the holographic principle. It therefore owes its existence, at least in part, to the comparatively humble idea that the entropy of a black hole scales like the horizon area \eqref{eq:td1}. In turn, AdS/CFT has inspired a number of generalizations, extensions, and applications which may be considered descendants of black hole thermodynamics. Some early examples during the period 1993-2003 include duals of confining field theories with $\mathcal{N}=1$ supersymmetry \cite{Polchinski:2000uf}, the dS/CFT correspondence relating quantum gravity on de Sitter space to a Euclidean CFT \cite{Strominger:2001pn, Strominger:2001gp}, proposed duals of $O(N)$ vector models in terms of higher spin gauge theories \cite{Klebanov:2002ja}, and even applications of gauge/gravity duality techniques to calculations in inflationary cosmology \cite{Maldacena:2002vr, Larsen:2002et, Larsen:2003pf}.

The topics discussed in this section represent major achievements during the period 1993-2003, but they were certainly not the only interesting developments during that time. For instance, in 1995 Jacobson was able to extract, under certain assumptions, the Einstein equations from horizon thermodynamics \cite{Jacobson:1995ab}. This result inspired a fair amount of subsequent work, especially in recent years  \cite{Padmanabhan:2009vy, Verlinde:2010hp}.

\section*{2003--2013}


The previous decade saw great progress in microscopic state counting, and the emergence of holography as an important and perhaps fundamental property of quantum gravity. In recent years there has been a focus on applications and generalizations of AdS/CFT, efforts to identify the gravitational states associated with a black hole, and attempts to comprehensively resolve the information paradox. Some new problems have arisen, but developments that touch on two or more of these issues suggest a convergence towards a deeper understanding of quantum gravity and black hole thermodynamics.

\runinhead{Tests and applications of AdS/CFT?}
Early tests of AdS/CFT spawned a number of further checks that probed different regimes of the correspondence. For instance, methods known from integrable systems, such as the thermodynamic Bethe Ansatz, allowed to check aspects of AdS/CFT beyond perturbation theory (in particular, for arbitrary values of the 't~Hooft coupling constant $\lambda$); see \cite{Beisert:2010jr} for a review. As the correspondence matured a number of new applications were uncovered. An emblematic example is the prediction of 
the ratio of shear viscosity $\eta$ to entropy density $s$ in the infinite coupling limit \cite{Policastro:2001yc,Kovtun:2004de}. 
\begin{equation}
\frac{\eta}{s} = \frac{\hbar}{4\pi k_B}
 \label{eq:td51}
\end{equation}
In relativistic heavy ion collisions the same order of magnitude was observed for $\eta/s$ (see \cite{Romatschke:2007mq}), which inspired both phenomenologists and theoreticians to apply AdS/CFT methods to the description of relativistic plasmas, see e.g.~\cite{Teaney:2009qa,CasalderreySolana:2011us,Shuryak:2011aa,DeWolfe:2013cua} for reviews. 
The key feature of the $\eta/s$ story is that a complicated calculation on the field theory side---determining the shear viscosity for a strongly coupled plasma---is mapped to a problem on the gravity side that is suitable for a bright PhD student. Indeed, Damour provided a comparable calculation in his PhD thesis already in 1979 \cite{PhD:Damour}.

\runinhead{Gauge/gravity correspondences.}
The past decade has seen numerous further attempts to phenomenologically apply ideas from the AdS/CFT correspondence to more general settings. These `gauge/gravity' correspondences began with deformations of AdS/CFT, but were soon extended to conjectured dualities between theories that bear little resemblance to asymptotically AdS gravity or ${\cal N}=4$ Super-Yang--Mills.
As above, the idea is to map complicated (strong coupling) problems on the gauge theory side to fairly simple problems on the gravity side. Examples include condensed matter applications such as cold atoms \cite{Son:2008ye,Balasubramanian:2008dm,Adams:2008wt}, Lifshitz fixed points with non-relativistic scaling symmetries \cite{Kachru:2008yh}, superfluids/superconductors \cite{Gubser:2008px,Hartnoll:2008vx,Hartnoll:2008kx}, non-Fermi liquids/strange metals \cite{Hartnoll:2009ns,Liu:2009dm,Faulkner:2010zz,Faulkner:2011tm} and the gravity/fluid correspondence \cite{Iqbal:2008by,Bredberg:2010ky,Compere:2011dx,Bredberg:2011jq,Compere:2012mt} (based on the membrane paradigm \cite{Thorne:1986iy}). Some applications of proposed gauge/gravity dualities to condensed matter systems are reviewed in \cite{McGreevy:2009xe,Sachdev:2010uj,Hartnoll:2011fn,Sachdev:2011wg,Iqbal:2011ae,Son:2012zz}.


\runinhead{Limits of holography.}
We discuss now in a bit more detail some extensions of the AdS/CFT correspondence that are more in line with the main topic of our review. An interesting theoretically motivated question to ask is, how general is holography? Originally, the holographic principle was motivated by avoidance of information loss and preservation of unitarity, but the way the AdS/CFT correspondence works makes it plausible that it could also apply to systems that are non-unitary. Moreover, if the holographic principle is a true statement about Nature then it should be realized in settings other than AdS, such as asymptotically flat or accelerating Friedmann--Lemaitre--Robertson--Walker spacetimes.
Finally, it is interesting to ask whether there are theories apart from string theory that permit a holographic description. A conclusive answer to this question would be an important achievement. If affirmative, then such theories might provide novel playgrounds for theoretical considerations about holography as well as new applications along the lines of AdS/CFT. If negative, we would have established a direct link between holography and string theory, i.e., holography would necessarily imply string theory. 

\runinhead{Non-unitary holography.}
Partly for simplicity and partly because there were many developments in the past decade, we restrict ourselves mostly to 3-dimensional theories of gravity in order to address the issues raised in the previous paragraph. Let us start with the question to what extent holography could apply to non-unitary theories. This question is somewhat delicate, because non-unitarity is often associated with some sickness of the theory. However, there are also systems that exhibit non-unitarity in a `controlled' way. This includes, for instance, open quantum systems and systems with quenched disorder. In a story with several interesting twists, it appears that TMG \eqref{eq:td35} at the critical point $\mu\ell=1$ corresponds to a log CFT, as suggested first in \cite{Grumiller:2008qz}. Log CFTs are specific non-unitary CFTs where two or more operators have degenerate scaling dimensions and the Hamiltonian acquires a Jordan block structure \cite{Flohr:2001zs,Gaberdiel:2001tr,lcft}. They are used, for example, in the description of systems with quenched disorder. A key element on the gravity side is the emergence of log modes \cite{Grumiller:2008qz}
\begin{equation}
\psi^{\textrm{\tiny log}}_{\alpha\beta} = \lim_{\varepsilon\to 0}\frac{\psi^M_{\alpha\beta}-\psi^L_{\alpha\beta}}{\varepsilon} = -2(it+\ln\cosh\rho)\,\psi^L_{\alpha\beta}
 \label{eq:td54}
\end{equation}
as linearized perturbation on the AdS background. The middle equation indicates the degeneration of the massive graviton modes $\psi^M$ with the `left-moving boundary graviton' modes $\psi^L$ (specific Einstein-gravity modes at linearized level). The latter are eigenstates of the Hamiltonian, $H\psi^L=i\partial_t \psi^L=h\psi^L$, while the former are not: $H\psi^{\textrm{\tiny log}}=i\partial_t \psi^{\textrm{\tiny log}}=h\psi^{\textrm{\tiny log}}+2\psi^L$. These two equations make manifest the Jordan block structure of the Hamiltonian $H$ when acting on the pair $\psi^{\textrm{\tiny log}}, \psi^L$. See \cite{Grumiller:2013at} for a full account of various checks, generalizations and possible applications of the AdS/log CFT correspondence. Thus, it seems that it is possible to extend the holographic principle to theories that exhibit non-unitarity in a controlled way.

\runinhead{Flat space holography.}
There was some progress on extracting features of the flat space S-matrix from AdS/CFT correlators, see e.g.~\cite{Susskind:1998vk,Polchinski:1999ry,Giddings:1999jq,Gary:2009ae,Gary:2009mi}, but it is still fair to say that efforts at flat-space holography have not met with a great deal of success in dimension four and above. In three dimensions one can essentially repeat the Brown--Henneaux construction, which was done by Barnich and Compere \cite{Barnich:2006av}. The asymptotic symmetry algebra was found to be the Bondi--van~der~Burg--Metzner--Sachs (BMS) algebra \cite{Bondi:1962,Sachs:1962} in three dimensions, which arises also as the ultra-relativistic contraction (or large AdS radius limit) of the two-dimensional conformal algebra. These algebras are also known as Galilean conformal algebras \cite{Bagchi:2009my,Bagchi:2009pe}, which led to the notion of a `BMS/GCA correspondence' \cite{Bagchi:2010zz}. A specific proposal for a flat space/CFT correspondence is flat space chiral gravity (TMG \eqref{eq:td35} in the limit $\ell\to\infty$ and $G\to\infty$, with $\mu G$ kept finite), which is conjectured to be dual to a chiral half of a CFT \cite{Bagchi:2012yk}. In particular, for central charge $c=24$ the conjectured CFT is a chiral half of the monster CFT (proposed originally by Witten in the context of Einstein gravity \cite{Witten:2007kt}), exactly like in the chiral gravity proposal by Li, Song and Strominger \cite{Li:2008dq}. Its partition function is given by the $J$-function \cite{Witten:2007kt,Maloney:2009ck} and due to chirality depends solely on the `left-moving' modular parameter $q$.
\begin{equation}
Z(q)=J(q)=\frac{1}{q} + 196884 \,q + {\cal O}(q^2)
\label{eq:td52}
\end{equation}
The number $196884$ is interpreted as one Virasoro descendant of the vacuum plus $196883$ primary fields corresponding to flat space cosmology horizon microstates (see below). The flat space chiral gravity quantum entropy $S=\ln{196883}\approx 12.2$ differs only by about $3\%$ from the semi-classical Bekenstein--Hawking result $S_{\textrm{\tiny BH}}=4\pi\approx 12.6$ (in suitable units). For quantum gravity applications the (flat space) chiral gravity situation seems optimal: there are quantum corrections that are not completely negligible (of the order of a few percent), but the theory is not ``ultra-quantum'' so that geometric notions associated with the semi-classical limit, like black hole horizons, can still be discussed meaningfully. Flat space cosmologies \cite{Cornalba:2002fi,Cornalba:2003kd} are the flat space analog of BTZ black holes \eqref{eq:td37} and permit a microstate counting similar to AdS \eqref{eq:td39}, see \cite{Barnich:2012xq,Bagchi:2012xr}. They are subject to a Hawking--Page like phase transition \cite{Bagchi:2013lma} so that at least in three dimensions cosmic evolution can be generated by heating (and gently stirring) flat space. For further aspects of flat space holography see e.g.~\cite{Barnich:2010eb,Barnich:2011mi,Barnich:2012aw,Barnich:2012rz,Bagchi:2012cy,Barnich:2013yka,Barnich:2013axa,Bagchi:2013qva,Costa:2013vza}.

\runinhead{Higher spin holography.} Remarkably, AdS spacetimes permit interacting massless particles with spin greater than two \cite{Fradkin:1986qy,Fradkin:1987ks,Vasiliev:1990en,Vasiliev:2003ev}. These `higher spin' theories could be relevant for the holographic description of certain sectors of large $N$ gauge theories. In particular, Klebanov and Polyakov proposed that a particular Vasiliev-type higher spin theory on AdS$_4$ might be exactly dual to the $O(N)$ vector model (at large $N$) in three dimensions. This conjecture triggered an intensive study of the subject with impressive achievements \cite{Giombi:2009wh,Giombi:2010vg,Koch:2010cy,Giombi:2011ya}. 
An interesting technical aspect of higher spin holography is that it provides a weak/weak duality and therefore allows to test holography with high precision (the other side of the coin is that higher spin holography is of less practical use than AdS/CFT, since strong/weak dualities can map hard calculations to simple ones).
Coming back to three bulk dimensions, Henneaux and Rey (and independently Campoleoni, Fredenhagen, Pfenninger and Theisen) generalized the Brown--Henneaux analysis to higher spin theories with AdS boundary conditions \cite{Henneaux:2010xg,Campoleoni:2010zq}, and a few months later Gaberdiel and Gopakumar proposed a correspondence between Vasiliev-type higher spin gravity and minimal model CFTs \cite{Gaberdiel:2010pz,Gaberdiel:2011zw,Gaberdiel:2012uj}. Some selected papers and reviews are \cite{Fotopoulos:2008ka,Sagnotti:2010at,Bekaert:2010hw,Campoleoni:2011hg,Ammon:2011nk,Anninos:2011ui,Maldacena:2011jn,Ammon:2012wc,Vasiliev:2012vf,Maldacena:2012sf}. Recently, the topics of flat space and higher spin holography were combined \cite{Afshar:2013vka,Gonzalez:2013oaa}, in the spirit of non-AdS holography for three dimensional higher spin gravity \cite{Gary:2012ms}. The main observation is that unlike the spin-2 case, higher spin theories allow for many different backgrounds, including Lobachevsky, Lifshitz, Schr\"odinger and warped AdS besides more common backgrounds such as AdS or flat space, without the addition of matter fields.

\newcommand{\tr}{\textrm{tr}}  
\runinhead{Holographic entanglement entropy.} Entanglement entropy is an entanglement measure for bipartite pure states $|\Psi\rangle$ and is defined as the von Neumann entropy 
\begin{equation}
S_A = -\tr\rho_A\ln\rho_A = -\lim_{n\to 1}\,\frac{\extd}{\extd n}\, \tr\rho_A^n
\label{eq:ee}
\end{equation}
associated with the reduced density matrix $\rho_A=\tr_B |\Psi\rangle\langle\Psi|$ of a subsystem $A$, where the total system is divided into two subsystems $A$ and $B$, see e.g.~\cite{Nielsen:2000}. 
For the present context one can think of $A$ ($B$) as the exterior (interior) of a black hole.
Then $S_A$ can be thought of as the entropy for an observer who has access only to the black hole exterior.
The fact that entanglement entropy obeys an area law led to the suggestion \cite{Sorkin:1985bu, Bombelli:1986rw,Srednicki:1993im} that Bekenstein--Hawking entropy could be interpreted as entanglement entropy (see \cite{Solodukhin:2011gn} for a review).

Entanglement entropy has found many applications in quantum systems, see e.g.~\cite{Vidal:2002rm,Calabrese:2004eu,Kitaev:2005dm,Levin:2006zz}, but is not easy to calculate in interacting quantum field theories in dimension greater than two [sometimes the so-called replica trick can be used, which exploits the second equality in \eqref{eq:ee}]. The holographic entanglement entropy proposal by Ryu and Takayanagi \cite{Ryu:2006bv,Ryu:2006ef} applies holographic ideas to map the difficult calculation of entanglement entropy on the field theory side to an elementary calculation of minimal surfaces on the gravity side. This proposal has passed several tests by successfully reproducing the entanglement entropy in well-understood cases, see \cite{Nishioka:2009un} for a review. Taking the proposal for granted it can then be applied to situations in which no other method exists (currently) to determine entanglement entropy. Thus, holographic entanglement entropy, which provides a link between black hole thermodynamics and quantum information,\footnote{%
We mention in the conclusions that this link is likely to grow stronger in the future. Besides the numerous recent papers on holographic entanglement entropy, some selected papers that also provide such links are \cite{Hayden:2007cs,Harlow:2013tf} and references therein.
} is another example of the utility of weak/strong dualities like AdS/CFT.

\runinhead{Geometry of black hole thermodynamics and cosmological constant as state parameter.}
Over the last decade most of the work inspired by black hole thermodynamics focused on holography, AdS/CFT, and related issues. But there were also some interesting purely thermodynamical developments. For example, the geometry of black hole thermodynamics was investigated in numerous papers, see for instance \cite{Aman:2003ug,Arcioni:2004ww,Aman:2005xk,Shen:2005nu,Sarkar:2006tg,Alvarez:2008wa,Ruppeiner:2008kd}. The basic idea goes back to Ruppeiner \cite{Ruppeiner:1979,RevModPhys.67.605,RevModPhys.68.313}, namely to associate the Hessian of the entropy (with respect to some state space variables $x^i$) with a metric, $g_{ij}=-\partial_i\partial_j S(x^k)$, whose geometric properties are related to the thermodynamics of the system.\par
 Another example is the recent revival of the idea to treat the cosmological constant as a state parameter. This concept goes back to the germinal work of \cite{Henneaux:1984ji,Henneaux:1985tv}. In order to treat the cosmological constant as a state variable, $\Lambda$ is introduced as a constant of integration by coupling a four form field strength to gravity. The value of the cosmological constant can change by spontaneous nucleation of membranes that act as sources for the four form \cite{Brown:1987dd,Brown:1988kg,Bousso:2000xa}, or by thermal decay together with the creation of a black hole \cite{Gomberoff:2003zh}. These results motivate the study of black hole thermodynamics in AdS, in a phase space extended by $\Lambda$ and its conjugate variable $\Theta$, the negative of which turns out to be a suitable ``thermodynamic" definition for the volume of a black hole \cite{Caldarelli:1999xj,Kastor:2009wy,Cvetic:2010jb,Dolan:2010ha,Dolan:2011xt,Kubiznak:2012wp,Dolan:2013dga,Dolan:2013ft}. See \cite{Altamirano:2014tva} for a review.

\runinhead{Kerr/CFT.}
The counting of black hole microstates pioneered by Strominger and Vafa in the previous decade initially was restricted to simple but astrophysically irrelevant black holes. In the decade discussed in this section a similar counting was applied to Kerr black holes, which established the `Kerr/CFT' correspondence, see e.g.~\cite{Guica:2008mu,Lu:2008jk,Azeyanagi:2008dk,Bredberg:2009pv,Cvetic:2009jn,Castro:2009jf,Dias:2009ex,Amsel:2009ev,Castro:2010fd,Guica:2010ej,Compere:2012jk}. Particularly the early papers were based on the near horizon extremal Kerr (NHEK) metric constructed by Bardeen and Horowitz \cite{Bardeen:1999px}.
\begin{equation}
\extd s^2_{\textrm{\tiny NHEK}}=M^2(1+\cos^2\!\theta)\, \Big(\frac{-\extd\hat t^2+\extd\hat r^2}{\hat r^2}\\
+\frac{4\sin^2\!\theta}{(1+\cos^2\!\theta)^2}\big(\extd\hat\phi+\frac{\extd\hat t}{\hat r}\big)^2+\extd\theta^2\Big)
 \label{eq:td53}
\end{equation}
The line-element \eqref{eq:td53} is obtained from the Kerr geometry \eqref{eq:kerr} by rescaling $\hat t = \frac{\lambda t}{2M}$, $\hat r = \frac{\lambda M}{r-M}$, $\hat\phi=\phi-\frac{t}{2M}$ and taking the limit $\lambda\to 0$ while keeping $\hat t, \hat r, \hat\phi, \theta$ fixed. The entropy counted by CFT methods then matches the Bekenstein--Hawking result \eqref{eq:td1}.
\begin{equation}
 S_{\textrm{CFT}} = \frac{2\pi J}{\hbar} = \frac{A_h}{4G} = S_{\textrm{\tiny BH}}
\end{equation}
While astrophysical black holes are never exactly extremal (the Thorne bound on the dimensionless Kerr parameter is $a<0.998$ \cite{Thorne:1974ve}), some of them come very close to this bound. A possible example is GRS1915+105 whose dimensionless Kerr parameter appears to exceed $a \gtrsim 0.98$ \cite{McClintock:2006xd,McClintock:2009as} (however, see \cite{Fender:2010tk}).

\runinhead{Fuzzballs.} The various successes of counting black hole microstates all failed to answer an important question: what do the corresponding microstate geometries look like? The fuzzball proposal \cite{Lunin:2001jy} addresses this question in the context of string theory, stating that there should be ${\cal O}(e^S)$ horizonless and regular solutions that asymptote to the geometry of a given black hole, but differ from this geometry at the scale of the horizon, see also \cite{Lin:2004nb,Grant:2005qc}, and \cite{Mathur:2005zp,Bena:2007kg,Skenderis:2008qn,Mathur:2008nj} for reviews. The fuzzball proposal is motivated by the AdS/CFT correspondence as follows: for every state in the CFT counted by the Cardy formula there is a corresponding regular asymptotically AdS geometry. Each of these geometries encodes the vacuum expectation values of gauge invariant operators in that state through the standard AdS/CFT dictionary. These solutions can be stringy in the interior, though large classes of solutions have been identified that are well-described by the supergravity approximation. 

One of the main achievements claimed by proponents of the fuzzball program is a resolution of the information paradox. 
This can be traced back to a key property of the proposal, which is that quantum gravity effects in string theory can take place on scales much larger than the Planck scale due to `fractionation' \cite{Mathur:2009hf}. This results in significant modifications of Hawking radiation at wavelengths of order $G\,M$, allowing information to escape the `hole' and be recovered (in principle) by external observers. On the other hand, if the system is probed with some object of sufficiently high energy, $E\gg T$ (where $T$ is the Hawking temperature), then collective modes of the fuzzball are excited, which is well-approximated by a description in terms of an ensemble average over all fuzzballs. The latter reproduces the black hole geometry, so that the dynamics of sufficiently energetic objects over short timescales (like an astronaut falling into a black hole) are essentially the same as one would expect in a classical black hole geometry.  

\runinhead{Firewalls.} The information loss problem has resurfaced in the past few years through an ingenious gedankenexperiment set up by Almheiri, Marolf, Polchinski and Sully (AMPS) that highlighted their difficulty of reconciling black hole complementarity with the equivalence principle \cite{Almheiri:2012rt}. AMPS and several other authors argued that a possible resolution of this incompatibility results in an infalling observer encountering a `firewall' close to the horizon, see \cite{Almheiri:2013hfa,Marolf:2013dba} and references therein. The AMPS gedankenexperiment has engendered a lot of discussion and is an excellent demonstration that, at least collectively, the days of confusion regarding black hole thermodynamics and information loss are not over yet. There is of course a simple resolution of the apparent firewall paradox, but the margin is too small to include it here.

\section*{Conclusions and Future}


\newcommand{\numb}{\gamma} 

\runinhead{Log corrections to entropy.} We started our journey through the past fifty years of black hole thermodynamics with the Bekenstein--Hawking relation \eqref{eq:td1} and the statement that black hole thermodynamics provides non-trivial consistency checks for quantum gravity. We will end our review on a similar note, by going one step further than Bekenstein and Hawking. Namely, in the semi-classical approximation the area law obtains quantum corrections, which can be organized in an expansion in terms of $1/S_{\textrm{\tiny BH}}$ (the same kind of correction is obtained from subleading contributions to the Cardy formula \cite{Carlip:2000nv}).
\begin{equation}
S = S_{\textrm{\tiny BH}} + \numb_1\,\ln S_{\textrm{\tiny BH}} + \numb_2 + {\cal O}(1/S_{\textrm{\tiny BH}})
 \label{eq:td55}
\end{equation}
While the subleading terms and the ${\cal O}(1)$ term depend on the specific quantum gravity theory, the leading and first subleading term depend only on the classical limit of that theory and the validity of the semi-classical approximation. In other words, any theory of quantum gravity that is supposed to be equivalent to Einstein gravity in its semi-classical limit must not only reproduce the Bekenstein--Hawking law \eqref{eq:td1}, but also the same result for the numerical coefficient $\numb_1$ in front of the logarithmic correction term as obtained from perturbative (1-loop) quantization of Einstein gravity (with a given set of matter degrees of freedom --- the precise coefficient depends on the specific matter content). A recent summary of logarithmic corrections to Schwarzschild and other non-extremal black holes was provided by Sen \cite{Sen:2012dw}. He found that string theory calculations, whenever available, agree precisely with the semi-classical result. Interestingly, the simplest of all black holes, the Schwarzschild black hole, still presents a challenge: currently, string theory does not provide a prediction for $\numb_1$ of the Schwarzschild black hole.\footnote{Loop quantum gravity does provide such a prediction \cite{Kaul:1998xv,Kaul:2000kf}, and it disagrees with the semi-classical result.}

\runinhead{Future developments.} Predictions of future developments often serve as a source of amusement for future generations \cite{IBM}, but we will venture one as our closing statement. While a lot of our current understanding of black hole thermodynamics and quantum gravity was achieved through consistently applying Feynman's dictum ``everything is particle'' --- most prominently epitomized by the Hawking effect --- we predict that most of our future understanding will be achieved through consistently applying Wheeler's dictum  ``everything is information'' \cite{Wheeler}, like in the recent slogan ``ER = EPR'' \cite{Maldacena:2013xja} that emerged from the firewall discussions.

\begin{acknowledgement}
DG and RM thank their respective collaborators for numerous discussions on black hole thermodynamics in the past 15 years.
DG and JS were supported by the START project Y~435-N16 of the Austrian Science Fund (FWF) and the FWF projects I~952-N16 and I~1030-N27. 
\end{acknowledgement}


\begin{thebibliography}{100}

\bibitem{Carlip:2001wq}
S.~Carlip, ``{Quantum gravity: A progress report},'' {\em Rept. Prog. Phys.}
  {\bf 64} (2001) 885,
\href{http://www.arXiv.org/abs/arXiv:gr-qc/0108040}{{\tt arXiv:gr-qc/0108040}}.

\bibitem{Kerr:1963ud}
R.~P. Kerr, ``Gravitational field of a spinning mass as an example of
  algebraically special metrics,'' {\em Phys. Rev. Lett.} {\bf 11} (1963)
237--238.

\bibitem{Bekenstein:1972tm}
J.~Bekenstein, ``{Black holes and the second law},'' {\em Lett.Nuovo Cim.} {\bf
  4} (1972)
737--740.

\bibitem{Bardeen:1973gs}
J.~M. Bardeen, B.~Carter, and S.~Hawking, ``{The Four laws of black hole
  mechanics},'' {\em Commun.Math.Phys.} {\bf 31} (1973)
161--170.

\bibitem{tolman1987relativity}
R.~Tolman, {\em Relativity, Thermodynamics, and Cosmology}.
\newblock Dover Books on Physics Series. Dover Publications, 1987.

\bibitem{Oppenheimer:1939ne}
J.~Oppenheimer and G.~Volkoff, ``{On Massive neutron cores},'' {\em Phys.Rev.}
  {\bf 55} (1939)
374--381.

\bibitem{Tolman:Static}
R.~C. Tolman, ``Static solutions of einstein's field equations for spheres of
  fluid,'' {\em Phys. Rev.} {\bf 55} (Feb, 1939) 364--373.

\bibitem{Zel:1961}
Y.~B. Zel'dovich {\em Zh. Eksp. Teoret. Fiz.} {\bf 41} (1961) 1609.

\bibitem{Bondi:Massive}
H.~Bondi, ``{Massive spheres in general relativity},'' {\em Proc.Roy.Soc.Lond.}
  {\bf A281} (1964)
303--317.

\bibitem{Sorkin:1981wd}
R.~D. Sorkin, R.~M. Wald, and Z.~J. Zhang, ``{Entropy of selfgravitating
  radiation},'' {\em Gen.Rel.Grav.} {\bf 13} (1981)
1127--1146.

\bibitem{Newman:1965yu}
E.~T. Newman, R.~Couch, K.~Chinnapared, A.~Exton, A.~Prakash, {\em et al.},
  ``{Metric of a Rotating, Charged Mass},'' {\em J.Math.Phys.} {\bf 6} (1965)
918--919.

\bibitem{Ginzburg:1965uf}
V.~Ginzburg and L.~Ozernoi {\em Sov.Phys. JETP 20} (1965) 689.

\bibitem{Doroshkevich:1966ru}
A.~Doroshkevich, Y.~Zel'dovich, and I.~Novikov, ``Gravitational collapse of
  nonsymmetric and rotating masses,'' {\em JETP 49} (1965).

\bibitem{Israel:1967wq}
W.~Israel, ``{Event horizons in static vacuum space-times},'' {\em Phys.Rev.}
  {\bf 164} (1967)
1776--1779.

\bibitem{Israel:1967za}
W.~Israel, ``{Event horizons in static electrovac space-times},'' {\em
  Commun.Math.Phys.} {\bf 8} (1968)
245--260.

\bibitem{Carter:1971zc}
B.~Carter, ``{Axisymmetric Black Hole Has Only Two Degrees of Freedom},'' {\em
  Phys.Rev.Lett.} {\bf 26} (1971)
331--333.

\bibitem{Robinson:1975bv}
D.~Robinson, ``{Uniqueness of the Kerr black hole},'' {\em Phys.Rev.Lett.} {\bf
  34} (1975)
905--906.

\bibitem{Mazur:1982db}
P.~Mazur, ``{Proof of Uniqueness of the Kerr-Newman Black Hole Solution},''
  {\em J.Phys.} {\bf A15} (1982)
3173--3180.

\bibitem{Bunting:1983bh}
G.~Bunting, {\em Proof of the Uniqueness Conjecture for Black Holes}.
\newblock PhD thesis, Univ. New England, Armadale N.S.W., 1983.

\bibitem{Chrusciel:2008js}
P.~T. Chrusciel and J.~Lopes~Costa, ``{On uniqueness of stationary vacuum black
  holes},'' {\em Asterisque} {\bf 321} (2008) 195--265,
\href{http://www.arXiv.org/abs/0806.0016}{{\tt 0806.0016}}.

\bibitem{Chrusciel:2012jk}
P.~T. Chrusciel, J.~L. Costa, and M.~Heusler, ``{Stationary Black Holes:
  Uniqueness and Beyond},'' {\em Living Rev.Rel.} {\bf 15} (2012) 7,
\href{http://www.arXiv.org/abs/1205.6112}{{\tt 1205.6112}}.

\bibitem{Misner:1973qy}
C.~W. Misner, K.~Thorne, and J.~Wheeler, {\em {Gravitation}}.
\newblock W.H. Freeman,
1973.
\newblock

\bibitem{Penrose:1969pc}
R.~Penrose, ``Gravitational collapse: {T}he role of general relativity,'' {\em
  Riv. Nuovo Cim.} {\bf 1} (1969)
252--276.

\bibitem{Christodoulou:1970wf}
D.~Christodoulou, ``{Reversible and irreversible transformations in black hole
  physics},'' {\em Phys.Rev.Lett.} {\bf 25} (1970)
1596--1597.

\bibitem{Christodoulou:1972kt}
D.~Christodoulou and R.~Ruffini, ``{Reversible transformations of a charged
  black hole},'' {\em Phys.Rev.} {\bf D4} (1971)
3552--3555.

\bibitem{Misner:1972kx}
C.~W. Misner, ``{Interpretation of gravitational-wave observations},'' {\em
  Phys.Rev.Lett.} {\bf 28} (1972)
994--997.

\bibitem{Press:1972zz}
W.~H. Press and S.~A. Teukolsky, ``{Floating Orbits, Superradiant Scattering
  and the Black-hole Bomb},'' {\em Nature} {\bf 238} (1972)
211--212.

\bibitem{Starobinsky:1972am}
A.~Starobinsky, ``Amplification of waves during reflection from a rotating
  black hole,'' {\em JETP} {\bf 64} (1972), no.~1, 28.

\bibitem{Zeldovich:1971am}
Y.~B. Zel'dovich, ``Amplification of cylindrical electromagnetic waves
  reflected from a rotating body,'' {\em JETP} {\bf 35} (1971), no.~6, 1085.

\bibitem{Zeldovich:1971le}
Y.~B. Zel'dovich, ``Generation of waves by a rotating body,'' {\em Sov.Phys.
  JETP Lett. 14} (1971) 180.

\bibitem{Carter:1968rr}
B.~Carter, ``{Global structure of the Kerr family of gravitational fields},''
  {\em Phys. Rev.} {\bf 174} (1968)
1559--1571.

\bibitem{Teukolsky:1973ha}
S.~A. Teukolsky, ``{Perturbations of a rotating black hole. 1. Fundamental
  equations for gravitational electromagnetic and neutrino field
  perturbations},'' {\em Astrophys.J.} {\bf 185} (1973)
635--647.

\bibitem{Teukolsky:1974yv}
S.~Teukolsky and W.~Press, ``{Perturbations of a rotating black hole. III -
  Interaction of the hole with gravitational and electromagnet ic radiation},''
  {\em Astrophys.J.} {\bf 193} (1974)
443--461.

\bibitem{Starobinsky:1973ch}
A.~Starobinsky and S.~Churilov, ``Amplification of electromagnetic and
  gravitational waves scattered by a rotating black hole,'' {\em JETP} {\bf 38}
  (1974), no.~1, 1.

\bibitem{Unruh:1973ne}
W.~Unruh, ``Separability of the neutrino equations in a kerr background,'' {\em
  Phys. Rev. Lett.} {\bf 31} (Nov, 1973) 1265--1267.

\bibitem{Hawking:1974sw}
S.~Hawking, ``{Particle Creation by Black Holes},'' {\em Commun.Math.Phys.}
  {\bf 43} (1975)
199--220.

\bibitem{Klein:1929zz}
O.~Klein, ``{Die Reflexion von Elektronen an einem Potentialsprung nach der
  relativistischen Dynamik von Dirac},'' {\em Z.Phys.} {\bf 53} (1929)
157.

\bibitem{Dombey:1999id}
N.~Dombey and A.~Calogeracos, ``{Seventy years of the Klein paradox},'' {\em
  Phys.Rept.} {\bf 315} (1999)
41--58.

\bibitem{Wald:1984rg}
R.~M. Wald, {\em {General Relativity}}.
\newblock The University of Chicago Press,
1984.
\newblock

\bibitem{Hawking:1974rv}
S.~W. Hawking, ``Black hole explosions,'' {\em Nature} {\bf 248} (1974)
30--31.

\bibitem{Hawking:1973uf}
S.~Hawking and G.~Ellis, {\em {The Large scale structure of space-time}}.
\newblock Cambridge University Press, Cambridge,
1973.
\newblock

\bibitem{Carter:1973lh}
B.~Carter, ``Black hole equilibrium states,'' in {\em Black Holes - Les astres
  occlus}.
\newblock Gordon and Breach Science Publishers, 1973.

\bibitem{Penrose:1968ar}
R.~Penrose, ``{Structure of space-time},'' in {\em Battelle Rencontres, 1967
  Lectures in Mathematics and Physics}, J.~W. C.M.~DeWitt, ed.
\newblock Benjamin, New York,
1968.
\newblock

\bibitem{Hawking:1971tu}
S.~W. Hawking, ``Gravitational radiation from colliding black holes,'' {\em
  Phys. Rev. Lett.} {\bf 26} (1971)
1344--1346.

\bibitem{Hawking:1971vc}
S.~Hawking, ``{Black holes in general relativity},'' {\em Commun.Math.Phys.}
  {\bf 25} (1972)
152--166.

\bibitem{Bekenstein:1973mi}
J.~Bekenstein, ``{Extraction of energy and charge from a black hole},'' {\em
  Phys.Rev.} {\bf D7} (1973)
949--953.

\bibitem{Wald:1972sz}
R.~M. Wald, ``{Gravitational spin interaction},'' {\em Phys.Rev.} {\bf D6}
  (1972)
406--413.

\bibitem{Smarr:1972kt}
L.~Smarr, ``{Mass formula for Kerr black holes},'' {\em Phys.Rev.Lett.} {\bf
  30} (1973)
71--73.

\bibitem{Wald:1994qf}
R.~M. Wald, {\em Quantum field theory in curved spacetime and black hole
  thermodynamics}.
\newblock University of Chicago Press, 1994.

\bibitem{Nernst:1912th}
W.~Nernst, ``Thermodynamik und spezifische w{\"a}rme,'' {\em Preuss. Akad.
  Wiss. Sitzungsberichte} {\bf 1} (1912) 134140.

\bibitem{Israel:1986tl}
W.~Israel, ``Third law of black-hole dynamics: A formulation and proof,'' {\em
  Phys. Rev. Lett.} {\bf 57} (Jul, 1986) 397--399.

\bibitem{Bekenstein:1973ur}
J.~D. Bekenstein, ``Black holes and entropy,'' {\em Phys. Rev.} {\bf D7} (1973)
2333--2346.

\bibitem{Grishchuk:1990bj}
L.~Grishchuk and Y.~Sidorov, ``{Squeezed quantum states of relic gravitons and
  primordial density fluctuations},'' {\em Phys.Rev.} {\bf D42} (1990)
3413--3421.

\bibitem{Wald:1975kc}
R.~M. Wald, ``{On Particle Creation by Black Holes},'' {\em Commun.Math.Phys.}
  {\bf 45} (1975)
9--34.

\bibitem{Parker:1975jm}
L.~Parker, ``{Probability Distribution of Particles Created by a Black Hole},''
  {\em Phys.Rev.} {\bf D12} (1975)
1519--1525.

\bibitem{Hawking:1976ra}
S.~W. Hawking, ``Breakdown of predictability in gravitational collapse,'' {\em
  Phys. Rev.} {\bf D14} (1976)
2460--2473.

\bibitem{Bekenstein:1977mv}
J.~Bekenstein and A.~Meisels, ``{Einstein a and B Coefficients for a Black
  Hole},'' {\em Phys.Rev.} {\bf D15} (1977)
2775--2781.

\bibitem{Panangaden:1977pc}
P.~Panangaden and R.~M. Wald, ``{Probability Distribution for Radiation from a
  Black Hole in the Presence of Incoming Radiation},'' {\em Phys.Rev.} {\bf
  D16} (1977)
929--932.

\bibitem{Boulware:1974dm}
D.~G. Boulware, ``{Quantum Field Theory in Schwarzschild and Rindler Spaces},''
  {\em Phys.Rev.} {\bf D11} (1975)
1404.

\bibitem{Gibbons:1978kq}
G.~Gibbons and M.~Perry, ``{Black Holes and Thermal Green's Functions},'' {\em
  Proc.Roy.Soc.Lond.} {\bf A358} (1978)
467--494.

\bibitem{Davies:1974th}
P.~Davies, ``{Scalar particle production in Schwarzschild and Rindler
  metrics},'' {\em J.Phys.} {\bf A8} (1975)
609--616.

\bibitem{DeWitt:1975ys}
B.~S. DeWitt, ``{Quantum Field Theory in Curved Space-Time},'' {\em Phys.Rept.}
  {\bf 19} (1975)
295--357.

\bibitem{Gerlach:1976ji}
U.~Gerlach, ``{The Mechanism of Black Body Radiation from an Incipient Black
  Hole},'' {\em Phys.Rev.} {\bf D14} (1976)
1479--1508.

\bibitem{Parikh:1999mf}
M.~K. Parikh and F.~Wilczek, ``Hawking radiation as tunneling,'' {\em Phys.
  Rev. Lett.} {\bf 85} (2000) 5042--5045,
\href{http://www.arXiv.org/abs/hep-th/9907001}{{\tt hep-th/9907001}}.

\bibitem{Christensen:1977jc}
S.~M. Christensen and S.~A. Fulling, ``Trace anomalies and the {H}awking
  effect,'' {\em Phys. Rev.} {\bf D15} (1977)
2088--2104.

\bibitem{Strominger:1994lh}
A.~Strominger, ``{Les Houches lectures on black holes},''
\href{http://www.arXiv.org/abs/hep-th/9501071}{{\tt hep-th/9501071}}.

\bibitem{Grumiller:2002nm}
D.~Grumiller, W.~Kummer, and D.~V. Vassilevich, ``Dilaton gravity in two
  dimensions,'' {\em Phys. Rept.} {\bf 369} (2002) 327--429,
\href{http://arXiv.org/abs/hep-th/0204253}{{\tt hep-th/0204253}}.

\bibitem{Robinson:2005pd}
S.~P. Robinson and F.~Wilczek, ``{A Relationship between Hawking radiation and
  gravitational anomalies},'' {\em Phys.Rev.Lett.} {\bf 95} (2005) 011303,
\href{http://www.arXiv.org/abs/gr-qc/0502074}{{\tt gr-qc/0502074}}.

\bibitem{Iso:2006ut}
S.~Iso, H.~Umetsu, and F.~Wilczek, ``{Anomalies, Hawking radiations and
  regularity in rotating black holes},'' {\em Phys.Rev.} {\bf D74} (2006)
  044017,
\href{http://www.arXiv.org/abs/hep-th/0606018}{{\tt hep-th/0606018}}.

\bibitem{Iso:2006wa}
S.~Iso, H.~Umetsu, and F.~Wilczek, ``{Hawking radiation from charged black
  holes via gauge and gravitational anomalies},'' {\em Phys.Rev.Lett.} {\bf 96}
  (2006) 151302,
\href{http://www.arXiv.org/abs/hep-th/0602146}{{\tt hep-th/0602146}}.

\bibitem{Gibbons:1978ac}
G.~Gibbons, S.~Hawking, and M.~Perry, ``{Path Integrals and the Indefiniteness
  of the Gravitational Action},'' {\em Nucl.Phys.} {\bf B138} (1978) 141.

\bibitem{Gibbons:1976ue}
G.~W. Gibbons and S.~W. Hawking, ``Action integrals and partition functions in
  quantum gravity,'' {\em Phys. Rev.} {\bf D15} (1977)
2752--2756.

\bibitem{Hawking:1978jz}
S.~Hawking, ``{Quantum Gravity and Path Integrals},'' {\em Phys.Rev.} {\bf D18}
  (1978)
1747--1753.

\bibitem{Gross:1982cv}
D.~J. Gross, M.~J. Perry, and L.~G. Yaffe, ``Instability of flat space at
  finite temperature,'' {\em Phys. Rev.} {\bf D25} (1982)
330--355.

\bibitem{Gibbons:1979xm}
G.~W. Gibbons and S.~W. Hawking, ``Classification of gravitational instanton
  symmetries,'' {\em Commun. Math. Phys.} {\bf 66} (1979)
291--310.

\bibitem{Gibbons:1979nf}
G.~Gibbons and M.~J. Perry, ``{New Gravitational Instantons and Their
  Interactions},'' {\em Phys.Rev.} {\bf D22} (1980)
313.

\bibitem{Gibbons:1978ji}
G.~Gibbons and M.~Perry, ``{Quantizing Gravitational Instantons},'' {\em
  Nucl.Phys.} {\bf B146} (1978)
90.

\bibitem{Zaumen:1974an}
W.~Zaumen, ``Upper bound on the electric charge of a black hole,'' {\em Nature}
  {\bf 247} (1974) 530.

\bibitem{Carter:1974yx}
B.~Carter, ``{Charge and particle conservation in black hole decay},'' {\em
  Phys.Rev.Lett.} {\bf 33} (1974)
558--561.

\bibitem{Gibbons:1975kk}
G.~Gibbons, ``{Vacuum Polarization and the Spontaneous Loss of Charge by Black
  Holes},'' {\em Commun.Math.Phys.} {\bf 44} (1975)
245--264.

\bibitem{Page:1976df}
D.~N. Page, ``{Particle Emission Rates from a Black Hole: Massless Particles
  from an Uncharged, Nonrotating Hole},'' {\em Phys.Rev.} {\bf D13} (1976)
198--206.

\bibitem{Page:1977um}
D.~N. Page, ``{Particle Emission Rates from a Black Hole. 3. Charged Leptons
  from a Nonrotating Hole},'' {\em Phys.Rev.} {\bf D16} (1977)
2402--2411.

\bibitem{Page:1976ki}
D.~N. Page, ``{Particle Emission Rates from a Black Hole. 2. Massless Particles
  from a Rotating Hole},'' {\em Phys.Rev.} {\bf D14} (1976)
3260--3273.

\bibitem{Unruh:1974bw}
W.~Unruh, ``{Second quantization in the Kerr metric},'' {\em Phys.Rev.} {\bf
  D10} (1974)
3194--3205.

\bibitem{Vilenkin:1979ui}
A.~Vilenkin, ``{MACROSCOPIC PARITY VIOLATING EFFECTS: NEUTRINO FLUXES FROM
  ROTATING BLACK HOLES AND IN ROTATING THERMAL RADIATION},'' {\em Phys.Rev.}
  {\bf D20} (1979)
1807--1812.

\bibitem{Leahy:1979xi}
D.~Leahy and W.~Unruh, ``{ANGULAR DEPENDENCE OF NEUTRINO EMISSION FROM ROTATING
  BLACK HOLES},'' {\em Phys.Rev.} {\bf D19} (1979)
3509--3515.

\bibitem{Fulling:1972md}
S.~A. Fulling, ``{Nonuniqueness of canonical field quantization in Riemannian
  space-time},'' {\em Phys.Rev.} {\bf D7} (1973)
2850--2862.

\bibitem{Unruh:1976db}
W.~G. Unruh, ``Notes on black hole evaporation,'' {\em Phys. Rev.} {\bf D14}
  (1976)
870.

\bibitem{Unruh:1983ac}
W.~G. Unruh and N.~Weiss, ``{Acceleration Radiation in Interacting Field
  Theories},'' {\em Phys.Rev.} {\bf D29} (1984)
1656.

\bibitem{Fulling:1987tp}
S.~Fulling and S.~Ruijsenaars, ``Temperature, periodicity and horizons,'' {\em
  Physics reports} {\bf 152} (1987), no.~3, 135--176.

\bibitem{Bisognano:1975ih}
J.~Bisognano and E.~Wichmann, ``{On the Duality Condition for a Hermitian
  Scalar Field},'' {\em J.Math.Phys.} {\bf 16} (1975)
985--1007.

\bibitem{Bisognano:1976za}
J.~Bisognano and E.~Wichmann, ``{On the Duality Condition for Quantum
  Fields},'' {\em J.Math.Phys.} {\bf 17} (1976)
303--321.

\bibitem{Sewell:1982zz}
G.~L. Sewell, ``{Quantum fields on manifolds: PCT and gravitationally induced
  thermal states},'' {\em Annals Phys.} {\bf 141} (1982)
201--224.

\bibitem{Unruh:1983ms}
W.~G. Unruh and R.~M. Wald, ``{What happens when an accelerating observer
  detects a Rindler particle},'' {\em Phys.Rev.} {\bf D29} (1984)
1047--1056.

\bibitem{Crispino:2007eb}
L.~C. Crispino, A.~Higuchi, and G.~E. Matsas, ``{The Unruh effect and its
  applications},'' {\em Rev.Mod.Phys.} {\bf 80} (2008) 787--838,
\href{http://www.arXiv.org/abs/0710.5373}{{\tt 0710.5373}}.

\bibitem{Bekenstein:1980jp}
J.~D. Bekenstein, ``{A Universal Upper Bound on the Entropy to Energy Ratio for
  Bounded Systems},'' {\em Phys.Rev.} {\bf D23} (1981)
287.

\bibitem{Unruh:1982ic}
W.~Unruh and R.~M. Wald, ``{Acceleration Radiation and Generalized Second Law
  of Thermodynamics},'' {\em Phys.Rev.} {\bf D25} (1982)
942--958.

\bibitem{Unruh:1983ir}
W.~Unruh and R.~M. Wald, ``{Entropy bounds, acceleration radiation, and the
  Generalized Second Law},'' {\em Phys.Rev.} {\bf D27} (1983)
2271--2276.

\bibitem{Hartle:1976tp}
J.~B. Hartle and S.~W. Hawking, ``Path integral derivation of black hole
  radiance,'' {\em Phys. Rev.} {\bf D13} (1976)
2188--2203.

\bibitem{Israel:1976ur}
W.~Israel, ``Thermo field dynamics of black holes,'' {\em Phys. Lett.} {\bf
  A57} (1976)
107--110.

\bibitem{Kay:1988mu}
B.~S. Kay and R.~M. Wald, ``{Theorems on the Uniqueness and Thermal Properties
  of Stationary, Nonsingular, Quasifree States on Space-Times with a Bifurcate
  Killing Horizon},'' {\em Phys.Rept.} {\bf 207} (1991)
49--136.

\bibitem{Unruh:1981ee}
W.~G. Unruh, ``Experimental black-hole evaporation?,'' {\em Phys. Rev. Lett.}
  {\bf 46} (May, 1981) 1351--1353.

\bibitem{Novello:2002qg}
M.~Novello, M.~Visser, and G.~Volovik, eds., {\em Artificial black holes}.
\newblock World Scientific, River Edge, USA, 2002.

\bibitem{Barcelo:2005fc}
C.~Barcelo, S.~Liberati, and M.~Visser, ``Analogue gravity,'' {\em Living Rev.
  Rel.} {\bf 8} (2005) 12,
\href{http://www.arXiv.org/abs/gr-qc/0505065}{{\tt gr-qc/0505065}}.

\bibitem{Ackermann:2009aa}
{\bf Fermi GBM/LAT Collaborations} Collaboration, M.~Ackermann {\em et al.},
  ``{A limit on the variation of the speed of light arising from quantum
  gravity effects},'' {\em Nature} {\bf 462} (2009) 331--334,
\href{http://www.arXiv.org/abs/0908.1832}{{\tt 0908.1832}}.

\bibitem{Jacobson:1991gr}
T.~Jacobson, ``{Black hole evaporation and ultrashort distances},'' {\em
  Phys.Rev.} {\bf D44} (1991)
1731--1739.

\bibitem{Unruh:1994je}
W.~Unruh, ``{Sonic analog of black holes and the effects of high frequencies on
  black hole evaporation},'' {\em Phys.Rev.} {\bf D51} (1995)
2827--2838.

\bibitem{Corley:1996ar}
S.~Corley and T.~Jacobson, ``{Hawking spectrum and high frequency
  dispersion},'' {\em Phys.Rev.} {\bf D54} (1996) 1568--1586,
\href{http://www.arXiv.org/abs/hep-th/9601073}{{\tt hep-th/9601073}}.

\bibitem{Jacobson:1993hn}
T.~Jacobson, ``{Black hole radiation in the presence of a short distance
  cutoff},'' {\em Phys.Rev.} {\bf D48} (1993) 728--741,
\href{http://www.arXiv.org/abs/hep-th/9303103}{{\tt hep-th/9303103}}.

\bibitem{Brout:1995wp}
R.~Brout, S.~Massar, R.~Parentani, and P.~Spindel, ``{Hawking radiation without
  transPlanckian frequencies},'' {\em Phys. Rev.} {\bf D52} (1995) 4559--4568,
\href{http://www.arXiv.org/abs/hep-th/9506121}{{\tt hep-th/9506121}}.

\bibitem{Visser:1997yu}
M.~Visser, ``{Hawking radiation without black hole entropy},'' {\em
  Phys.Rev.Lett.} {\bf 80} (1998) 3436--3439,
\href{http://www.arXiv.org/abs/gr-qc/9712016}{{\tt gr-qc/9712016}}.

\bibitem{Jacobson:1996zs}
T.~Jacobson, ``{On the origin of the outgoing black hole modes},'' {\em
  Phys.Rev.} {\bf D53} (1996) 7082--7088,
\href{http://www.arXiv.org/abs/hep-th/9601064}{{\tt hep-th/9601064}}.

\bibitem{Visser:1997ux}
M.~Visser, ``{Acoustic black holes: Horizons, ergospheres, and Hawking
  radiation},'' {\em Class. Quant. Grav.} {\bf 15} (1998) 1767--1791,
\href{http://www.arXiv.org/abs/gr-qc/9712010}{{\tt gr-qc/9712010}}.

\bibitem{Brown:1988}
J.~Brown, {\em Lower Dimensional Gravity}.
\newblock World Scientific, 1988.

\bibitem{Callan:1992rs}
C.~G. Callan, Jr., S.~B. Giddings, J.~A. Harvey, and A.~Strominger,
  ``Evanescent black holes,'' {\em Phys. Rev.} {\bf D45} (1992) 1005--1009,
\href{http://www.arXiv.org/abs/hep-th/9111056}{{\tt hep-th/9111056}}.

\bibitem{Russo:1992ax}
J.~G. Russo, L.~Susskind, and L.~Thorlacius, ``{The Endpoint of Hawking
  radiation},'' {\em Phys. Rev.} {\bf D46} (1992) 3444--3449,
\href{http://arXiv.org/abs/hep-th/9206070}{{\tt hep-th/9206070}}.

\bibitem{Susskind:1993if}
L.~Susskind, L.~Thorlacius, and J.~Uglum, ``{The Stretched horizon and black
  hole complementarity},'' {\em Phys.Rev.} {\bf D48} (1993) 3743--3761,
\href{http://www.arXiv.org/abs/hep-th/9306069}{{\tt hep-th/9306069}}.

\bibitem{Stephens:1994an}
C.~R. Stephens, G.~'t~Hooft, and B.~F. Whiting, ``Black hole evaporation
  without information loss,'' {\em Class. Quant. Grav.} {\bf 11} (1994)
  621--648,
\href{http://arXiv.org/abs/gr-qc/9310006}{{\tt gr-qc/9310006}}.

\bibitem{'tHooft:1984re}
G.~'t~Hooft, ``{On the Quantum Structure of a Black Hole},'' {\em Nucl.Phys.}
  {\bf B256} (1985)
727.

\bibitem{Jackiw:1984}
R.~Jackiw, ``{Liouville field theory: A two-dimensional model for gravity?},''
  in {\em Quantum Theory Of Gravity}, S.~Christensen, ed., pp.~403--420.
\newblock Adam Hilger, Bristol, 1984.

\bibitem{Teitelboim:1984}
C.~Teitelboim, ``{The {H}amiltonian structure of two-dimensional space-time and
  its relation with the conformal anomaly},'' in {\em Quantum Theory Of
  Gravity}, S.~Christensen, ed., pp.~327--344.
\newblock Adam Hilger, Bristol, 1984.

\bibitem{Isler:1989hq}
K.~Isler and C.~A. Trugenberger, ``A gauge theory of two-dimensional quantum
  gravity,'' {\em Phys. Rev. Lett.} {\bf 63} (1989)
834.

\bibitem{Chamseddine:1989yz}
A.~H. Chamseddine and D.~Wyler, ``Gauge theory of topological gravity in
  (1+1)-dimensions,'' {\em Phys. Lett.} {\bf B228} (1989)
75.

\bibitem{Cangemi:1992bj}
D.~Cangemi and R.~Jackiw, ``Gauge invariant formulations of lineal gravity,''
  {\em Phys. Rev. Lett.} {\bf 69} (1992) 233--236,
\href{http://arXiv.org/abs/hep-th/9203056}{{\tt hep-th/9203056}}.

\bibitem{Ikeda:1993aj}
N.~Ikeda and K.~I. Izawa, ``General form of dilaton gravity and nonlinear gauge
  theory,'' {\em Prog. Theor. Phys.} {\bf 90} (1993) 237--246,
\href{http://www.arXiv.org/abs/hep-th/9304012}{{\tt hep-th/9304012}}.

\bibitem{Ikeda:1993fh}
N.~Ikeda, ``{Two-dimensional gravity and nonlinear gauge theory},'' {\em Annals
  Phys.} {\bf 235} (1994) 435--464,
\href{http://www.arXiv.org/abs/hep-th/9312059}{{\tt hep-th/9312059}}.

\bibitem{Schaller:1994es}
P.~Schaller and T.~Strobl, ``Poisson structure induced (topological) field
  theories,'' {\em Mod. Phys. Lett.} {\bf A9} (1994) 3129--3136,
\href{http://arXiv.org/abs/hep-th/9405110}{{\tt hep-th/9405110}}.

\bibitem{Berger:1972pg}
B.~K. Berger, D.~M. Chitre, V.~E. Moncrief, and Y.~Nutku, ``Hamiltonian
  formulation of spherically symmetric gravitational fields,'' {\em Phys. Rev.}
  {\bf D5} (1972)
2467--2470.

\bibitem{Benguria:1977in}
R.~Benguria, P.~Cordero, and C.~Teitelboim, ``Aspects of the {H}amiltonian
  dynamics of interacting gravitational gauge and {H}iggs fields with
  applications to spherical symmetry,'' {\em Nucl. Phys.} {\bf B122} (1977)
61.

\bibitem{Thomi:1984na}
P.~Thomi, B.~Isaak, and P.~H{\'a}j{\'\i}{\v{c}}ek, ``Spherically symmetric
  systems of fields and black holes. 1. {D}efinition and properties of apparent
  horizon,'' {\em Phys. Rev.} {\bf D30} (1984)
1168.

\bibitem{Hajicek:1984mz}
P.~H{\'a}j{\'\i}{\v{c}}ek, ``Spherically symmetric systems of fields and black
  holes. 2. {A}pparent horizon in canonical formalism,'' {\em Phys. Rev.} {\bf
  D30} (1984)
1178.

\bibitem{Kuchar:1994zk}
K.~V. Kucha{\v{r}}, ``Geometrodynamics of {S}chwarzschild black holes,'' {\em
  Phys. Rev.} {\bf D50} (1994) 3961--3981,
\href{http://www.arXiv.org/abs/gr-qc/9403003}{{\tt gr-qc/9403003}}.

\bibitem{Soda:1993xc}
J.~Soda, ``{Hierarchical dimensional reduction and gluing geometries},'' {\em
  Prog.Theor.Phys.} {\bf 89} (1993)
1303--1310.

\bibitem{Emparan:2013xia}
R.~Emparan, D.~Grumiller, and K.~Tanabe, ``{Large D gravity and low D
  strings},'' {\em Phys.Rev.Lett.} {\bf 110} (2013) 251102,
\href{http://www.arXiv.org/abs/1303.1995}{{\tt 1303.1995}}.

\bibitem{Weinberg:1979}
S.~Weinberg in {\em General Relativity, an Einstein Centenary Survey},
  S.~Hawking and W.~Israel, eds.
\newblock Cambridge University Press, 1979.

\bibitem{Mann:1992ar}
R.~B. Mann and S.~F. Ross, ``{The D $\to$ 2 limit of general relativity},''
  {\em Class. Quant. Grav.} {\bf 10} (1993) 345--351,
\href{http://www.arXiv.org/abs/gr-qc/9208004}{{\tt gr-qc/9208004}}.

\bibitem{Grumiller:2007wb}
D.~Grumiller and R.~Jackiw, ``{Liouville gravity from Einstein gravity},'' in
  {\em {Recent Developments in Theoretical Physics}}, S.~Gosh and G.~Kar, eds.,
  pp.~331--343.
\newblock World Scientific, Singapore, 2010.
\newblock
\href{http://www.arXiv.org/abs/0712.3775}{{\tt 0712.3775}}.
\newblock

\bibitem{Ginsparg:1993is}
P.~Ginsparg and G.~W. Moore, ``Lectures on 2-d gravity and 2-d string theory,''
\href{http://arXiv.org/abs/hep-th/9304011}{{\tt hep-th/9304011}}.

\bibitem{Nakayama:2004vk}
Y.~Nakayama, ``{Liouville field theory: A decade after the revolution},'' {\em
  Int. J. Mod. Phys.} {\bf A19} (2004) 2771--2930,
\href{http://www.arXiv.org/abs/hep-th/0402009}{{\tt hep-th/0402009}}.

\bibitem{Katanaev:1986wk}
M.~O. Katanaev and I.~V. Volovich, ``String model with dynamical geometry and
  torsion,'' {\em Phys. Lett.} {\bf B175} (1986)
413--416.

\bibitem{Kummer:1992bg}
W.~Kummer and D.~J. Schwarz, ``{General analytic solution of R**2 gravity with
  dynamical torsion in two-dimensions},'' {\em Phys. Rev.} {\bf D45} (1992)
3628--3635.

\bibitem{Schaller:1992np}
P.~Schaller and T.~Strobl, ``{Canonical quantization of nonEinsteinian gravity
  and the problem of time},'' {\em Class. Quant. Grav.} {\bf 11} (1994)
  331--346,
\href{http://www.arXiv.org/abs/arXiv:hep-th/9211054}{{\tt
  arXiv:hep-th/9211054}}.

\bibitem{Katanaev:1995bh}
M.~O. Katanaev, W.~Kummer, and H.~Liebl, ``Geometric interpretation and
  classification of global solutions in generalized dilaton gravity,'' {\em
  Phys. Rev.} {\bf D53} (1996) 5609--5618,
\href{http://www.arXiv.org/abs/gr-qc/9511009}{{\tt gr-qc/9511009}}.

\bibitem{Callan:1985ia}
C.~G. Callan, Jr., E.~J. Martinec, M.~J. Perry, and D.~Friedan, ``Strings in
  background fields,'' {\em Nucl. Phys.} {\bf B262} (1985)
593.

\bibitem{Mandal:1991tz}
G.~Mandal, A.~M. Sengupta, and S.~R. Wadia, ``Classical solutions of
  two-dimensional string theory,'' {\em Mod. Phys. Lett.} {\bf A6} (1991)
1685--1692.

\bibitem{Elitzur:1991cb}
S.~Elitzur, A.~Forge, and E.~Rabinovici, ``Some global aspects of string
  compactifications,'' {\em Nucl. Phys.} {\bf B359} (1991)
581--610.

\bibitem{Witten:1991yr}
E.~Witten, ``On string theory and black holes,'' {\em Phys. Rev.} {\bf D44}
  (1991)
314--324.

\bibitem{Dijkgraaf:1992ba}
R.~Dijkgraaf, H.~Verlinde, and E.~Verlinde, ``String propagation in a black
  hole geometry,'' {\em Nucl. Phys.} {\bf B371} (1992)
269--314.

\bibitem{Klebanov:1991qa}
I.~R. Klebanov, ``{String theory in two-dimensions},''
\href{http://www.arXiv.org/abs/hep-th/9108019}{{\tt hep-th/9108019}}.

\bibitem{Strominger:1994tn}
A.~Strominger, ``Les {H}ouches lectures on black holes,''
  \href{http://www.arXiv.org/abs/arXiv:hep-th/9501071}{{\tt
  arXiv:hep-th/9501071}}.
Talk given at {NATO} Advanced Study Institute.

\bibitem{Grumiller:2006rc}
D.~Grumiller and R.~Meyer, ``Ramifications of lineland,'' {\em Turk. J. Phys.}
  {\bf 30} (2006) 349--378,
\href{http://www.arXiv.org/abs/hep-th/0604049}{{\tt hep-th/0604049}}.

\bibitem{Gegenberg:1994pv}
J.~Gegenberg, G.~Kunstatter, and D.~Louis-Martinez, ``Observables for
  two-dimensional black holes,'' {\em Phys. Rev.} {\bf D51} (1995) 1781--1786,
\href{http://www.arXiv.org/abs/gr-qc/9408015}{{\tt gr-qc/9408015}}.

\bibitem{Grumiller:2007ju}
D.~Grumiller and R.~McNees, ``Thermodynamics of black holes in two (and higher)
  dimensions,'' {\em JHEP} {\bf 04} (2007) 074,
\href{http://www.arXiv.org/abs/hep-th/0703230}{{\tt hep-th/0703230}}.

\bibitem{Wald:1993nt}
R.~M. Wald, ``{Black hole entropy is the Noether charge},'' {\em Phys.Rev.}
  {\bf D48} (1993) 3427--3431,
\href{http://www.arXiv.org/abs/gr-qc/9307038}{{\tt gr-qc/9307038}}.

\bibitem{Iyer:1994ys}
V.~Iyer and R.~M. Wald, ``{Some properties of Noether charge and a proposal for
  dynamical black hole entropy},'' {\em Phys.Rev.} {\bf D50} (1994) 846--864,
\href{http://www.arXiv.org/abs/gr-qc/9403028}{{\tt gr-qc/9403028}}.

\bibitem{York:1986it}
J.~W. York, Jr., ``{Black hole thermodynamics and the Euclidean Einstein
  action},'' {\em Phys. Rev.} {\bf D33} (1986)
2092--2099.

\bibitem{Hawking:1982dh}
S.~W. Hawking and D.~N. Page, ``Thermodynamics of black holes in anti-de
  {S}itter space,'' {\em Commun. Math. Phys.} {\bf 87} (1983)
577.

\bibitem{Deser:1982vy}
S.~Deser, R.~Jackiw, and S.~Templeton, ``Three-dimensional massive gauge
  theories,'' {\em Phys. Rev. Lett.} {\bf 48} (1982)
975--978.

\bibitem{Deser:1982wh}
S.~Deser, R.~Jackiw, and S.~Templeton, ``Topologically massive gauge
  theories,'' {\em Ann. Phys.} {\bf 140} (1982)
372--411.

\bibitem{Deser:1982a}
S.~Deser, R.~Jackiw, and S.~Templeton, ``Topologically massive gauge
  theories,'' {\em Erratum-ibid.} {\bf 185} (1988) 406.

\bibitem{Deser:1983tn}
S.~Deser, R.~Jackiw, and G.~'t~Hooft, ``{Three-Dimensional Einstein Gravity:
  Dynamics of Flat Space},'' {\em Annals Phys.} {\bf 152} (1984)
220.

\bibitem{Brown:1986nw}
J.~D. Brown and M.~Henneaux, ``{Central Charges in the Canonical Realization of
  Asymptotic Symmetries: An Example from Three-Dimensional Gravity},'' {\em
  Commun. Math. Phys.} {\bf 104} (1986)
207--226.

\bibitem{Banados:1992wn}
M.~Ba\~nados, C.~Teitelboim, and J.~Zanelli, ``The black hole in
  three-dimensional space-time,'' {\em Phys. Rev. Lett.} {\bf 69} (1992)
  1849--1851,
\href{http://www.arXiv.org/abs/hep-th/9204099}{{\tt hep-th/9204099}}.

\bibitem{Banados:1992gq}
M.~Ba\~nados, M.~Henneaux, C.~Teitelboim, and J.~Zanelli, ``Geometry of the
  (2+1) black hole,'' {\em Phys. Rev.} {\bf D48} (1993) 1506--1525,
\href{http://www.arXiv.org/abs/gr-qc/9302012}{{\tt gr-qc/9302012}}.

\bibitem{Cardy:1986ie}
J.~L. Cardy, ``Operator content of two-dimensional conformally invariant
  theories,'' {\em Nucl. Phys.} {\bf B270} (1986)
186--204.

\bibitem{Bloete:1986qm}
H.~W.~J. Bloete, J.~L. Cardy, and M.~P. Nightingale, ``Conformal invariance,
  the central charge, and universal finite size amplitudes at criticality,''
  {\em Phys. Rev. Lett.} {\bf 56} (1986)
742--745.

\bibitem{Strominger:1997eq}
A.~Strominger, ``Black hole entropy from near-horizon microstates,'' {\em JHEP}
  {\bf 02} (1998) 009,
\href{http://www.arXiv.org/abs/hep-th/9712251}{{\tt hep-th/9712251}}.

\bibitem{Carlip:1998wz}
S.~Carlip, ``Black hole entropy from conformal field theory in any dimension,''
  {\em Phys. Rev. Lett.} {\bf 82} (1999) 2828--2831,
\href{http://www.arXiv.org/abs/hep-th/9812013}{{\tt hep-th/9812013}}.

\bibitem{'tHooft:1993gx}
G.~'t~Hooft, ``{Dimensional reduction in quantum gravity},''
\href{http://www.arXiv.org/abs/gr-qc/9310026}{{\tt gr-qc/9310026}}.

\bibitem{Susskind:1993ws}
L.~Susskind, ``{Some speculations about black hole entropy in string theory},''
\href{http://www.arXiv.org/abs/hep-th/9309145}{{\tt hep-th/9309145}}.

\bibitem{Susskind:1994sm}
L.~Susskind and J.~Uglum, ``{Black hole entropy in canonical quantum gravity
  and superstring theory},'' {\em Phys.Rev.} {\bf D50} (1994) 2700--2711,
\href{http://www.arXiv.org/abs/hep-th/9401070}{{\tt hep-th/9401070}}.

\bibitem{Sen:1995in}
A.~Sen, ``{Extremal black holes and elementary string states},'' {\em
  Mod.Phys.Lett.} {\bf A10} (1995) 2081--2094,
\href{http://www.arXiv.org/abs/hep-th/9504147}{{\tt hep-th/9504147}}.

\bibitem{Strominger:1996sh}
A.~Strominger and C.~Vafa, ``{Microscopic Origin of the Bekenstein-Hawking
  Entropy},'' {\em Phys. Lett.} {\bf B379} (1996) 99--104,
\href{http://www.arXiv.org/abs/hep-th/9601029}{{\tt hep-th/9601029}}.

\bibitem{Maldacena:1996gb}
J.~M. Maldacena and A.~Strominger, ``{Statistical entropy of four-dimensional
  extremal black holes},'' {\em Phys.Rev.Lett.} {\bf 77} (1996) 428--429,
\href{http://www.arXiv.org/abs/hep-th/9603060}{{\tt hep-th/9603060}}.

\bibitem{Callan:1996dv}
C.~G. Callan and J.~M. Maldacena, ``{D-brane approach to black hole quantum
  mechanics},'' {\em Nucl.Phys.} {\bf B472} (1996) 591--610,
\href{http://www.arXiv.org/abs/hep-th/9602043}{{\tt hep-th/9602043}}.

\bibitem{Horowitz:1996fn}
G.~T. Horowitz and A.~Strominger, ``{Counting states of near extremal black
  holes},'' {\em Phys.Rev.Lett.} {\bf 77} (1996) 2368--2371,
\href{http://www.arXiv.org/abs/hep-th/9602051}{{\tt hep-th/9602051}}.

\bibitem{Emparan:2006it}
R.~Emparan and G.~T. Horowitz, ``{Microstates of a Neutral Black Hole in M
  Theory},'' {\em Phys.Rev.Lett.} {\bf 97} (2006) 141601,
\href{http://www.arXiv.org/abs/hep-th/0607023}{{\tt hep-th/0607023}}.

\bibitem{Skenderis:1999bs}
K.~Skenderis, ``{Black holes and branes in string theory},'' {\em Lect.Notes
  Phys.} {\bf 541} (2000) 325--364,
\href{http://www.arXiv.org/abs/hep-th/9901050}{{\tt hep-th/9901050}}.

\bibitem{Peet:2000hn}
A.~W. Peet, ``{TASI lectures on black holes in string theory},''
\href{http://www.arXiv.org/abs/hep-th/0008241}{{\tt hep-th/0008241}}.

\bibitem{Horowitz:2007xq}
G.~T. Horowitz and M.~M. Roberts, ``{Counting the Microstates of a Kerr Black
  Hole},'' {\em Phys.Rev.Lett.} {\bf 99} (2007) 221601,
\href{http://www.arXiv.org/abs/0708.1346}{{\tt 0708.1346}}.

\bibitem{Susskind:1994vu}
L.~Susskind, ``{The World as a hologram},'' {\em J.Math.Phys.} {\bf 36} (1995)
  6377--6396,
\href{http://www.arXiv.org/abs/hep-th/9409089}{{\tt hep-th/9409089}}.

\bibitem{Fischler:1998st}
W.~Fischler and L.~Susskind, ``{Holography and cosmology},''
\href{http://www.arXiv.org/abs/hep-th/9806039}{{\tt hep-th/9806039}}.

\bibitem{Easther:1999gk}
R.~Easther and D.~A. Lowe, ``{Holography, cosmology and the second law of
  thermodynamics},'' {\em Phys.Rev.Lett.} {\bf 82} (1999) 4967--4970,
\href{http://www.arXiv.org/abs/hep-th/9902088}{{\tt hep-th/9902088}}.

\bibitem{Bousso:1999xy}
R.~Bousso, ``{A Covariant entropy conjecture},'' {\em JHEP} {\bf 9907} (1999)
  004,
\href{http://www.arXiv.org/abs/hep-th/9905177}{{\tt hep-th/9905177}}.

\bibitem{Bousso:2002ju}
R.~Bousso, ``{The Holographic principle},'' {\em Rev.Mod.Phys.} {\bf 74} (2002)
  825--874,
\href{http://www.arXiv.org/abs/hep-th/0203101}{{\tt hep-th/0203101}}.

\bibitem{Klebanov:1997kc}
I.~R. Klebanov, ``{World volume approach to absorption by nondilatonic
  branes},'' {\em Nucl.Phys.} {\bf B496} (1997) 231--242,
\href{http://www.arXiv.org/abs/hep-th/9702076}{{\tt hep-th/9702076}}.

\bibitem{Gubser:1997yh}
S.~S. Gubser, I.~R. Klebanov, and A.~A. Tseytlin, ``{String theory and
  classical absorption by three-branes},'' {\em Nucl.Phys.} {\bf B499} (1997)
  217--240,
\href{http://www.arXiv.org/abs/hep-th/9703040}{{\tt hep-th/9703040}}.

\bibitem{Gubser:1997se}
S.~S. Gubser and I.~R. Klebanov, ``{Absorption by branes and Schwinger terms in
  the world volume theory},'' {\em Phys.Lett.} {\bf B413} (1997) 41--48,
\href{http://www.arXiv.org/abs/hep-th/9708005}{{\tt hep-th/9708005}}.

\bibitem{Witten:1995im}
E.~Witten, ``{Bound states of strings and p-branes},'' {\em Nucl.Phys.} {\bf
  B460} (1996) 335--350,
\href{http://www.arXiv.org/abs/hep-th/9510135}{{\tt hep-th/9510135}}.

\bibitem{Maldacena:1997re}
J.~M. Maldacena, ``{The large N limit of superconformal field theories and
  supergravity},'' {\em Adv. Theor. Math. Phys.} {\bf 2} (1998) 231--252,
\href{http://www.arXiv.org/abs/hep-th/9711200}{{\tt hep-th/9711200}}.

\bibitem{Beisert:2010jr}
N.~Beisert, C.~Ahn, L.~F. Alday, Z.~Bajnok, J.~M. Drummond, {\em et al.},
  ``{Review of AdS/CFT Integrability: An Overview},'' {\em Lett.Math.Phys.}
  {\bf 99} (2012) 3--32,
\href{http://www.arXiv.org/abs/1012.3982}{{\tt 1012.3982}}.

\bibitem{Witten:1998zw}
E.~Witten, ``{Anti-de Sitter space, thermal phase transition, and confinement
  in gauge theories},'' {\em Adv. Theor. Math. Phys.} {\bf 2} (1998) 505--532,
\href{http://www.arXiv.org/abs/hep-th/9803131}{{\tt hep-th/9803131}}.

\bibitem{Birmingham:1998jt}
D.~Birmingham, I.~Sachs, and S.~Sen, ``Entropy of three-dimensional black holes
  in string theory,'' {\em Phys. Lett.} {\bf B424} (1998) 275--280,
\href{http://www.arXiv.org/abs/hep-th/9801019}{{\tt hep-th/9801019}}.

\bibitem{Mathur:2009hf}
S.~D. Mathur, ``{The Information paradox: A Pedagogical introduction},'' {\em
  Class.Quant.Grav.} {\bf 26} (2009) 224001,
\href{http://www.arXiv.org/abs/0909.1038}{{\tt 0909.1038}}.

\bibitem{Polchinski:2000uf}
J.~Polchinski and M.~J. Strassler, ``{The String dual of a confining
  four-dimensional gauge theory},''
\href{http://www.arXiv.org/abs/hep-th/0003136}{{\tt hep-th/0003136}}.

\bibitem{Strominger:2001pn}
A.~Strominger, ``{The dS / CFT correspondence},'' {\em JHEP} {\bf 0110} (2001)
  034,
\href{http://www.arXiv.org/abs/hep-th/0106113}{{\tt hep-th/0106113}}.

\bibitem{Strominger:2001gp}
A.~Strominger, ``{Inflation and the dS / CFT correspondence},'' {\em JHEP} {\bf
  0111} (2001) 049,
\href{http://www.arXiv.org/abs/hep-th/0110087}{{\tt hep-th/0110087}}.

\bibitem{Klebanov:2002ja}
I.~Klebanov and A.~Polyakov, ``{AdS dual of the critical O(N) vector model},''
  {\em Phys.Lett.} {\bf B550} (2002) 213--219,
  \href{http://www.arXiv.org/abs/hep-th/0210114}{{\tt hep-th/0210114}}.

\bibitem{Maldacena:2002vr}
J.~M. Maldacena, ``{Non-Gaussian features of primordial fluctuations in single
  field inflationary models},'' {\em JHEP} {\bf 0305} (2003) 013,
\href{http://www.arXiv.org/abs/astro-ph/0210603}{{\tt astro-ph/0210603}}.

\bibitem{Larsen:2002et}
F.~Larsen, J.~P. van~der Schaar, and R.~G. Leigh, ``{De Sitter holography and
  the cosmic microwave background},'' {\em JHEP} {\bf 0204} (2002) 047,
\href{http://www.arXiv.org/abs/hep-th/0202127}{{\tt hep-th/0202127}}.

\bibitem{Larsen:2003pf}
F.~Larsen and R.~McNees, ``{Inflation and de Sitter holography},'' {\em JHEP}
  {\bf 0307} (2003) 051,
\href{http://www.arXiv.org/abs/hep-th/0307026}{{\tt hep-th/0307026}}.

\bibitem{Jacobson:1995ab}
T.~Jacobson, ``{Thermodynamics of space-time: The Einstein equation of
  state},'' {\em Phys.Rev.Lett.} {\bf 75} (1995) 1260--1263,
\href{http://www.arXiv.org/abs/gr-qc/9504004}{{\tt gr-qc/9504004}}.

\bibitem{Padmanabhan:2009vy}
T.~Padmanabhan, ``{Thermodynamical Aspects of Gravity: New insights},'' {\em
  Rept.Prog.Phys.} {\bf 73} (2010) 046901,
\href{http://www.arXiv.org/abs/0911.5004}{{\tt 0911.5004}}.

\bibitem{Verlinde:2010hp}
E.~P. Verlinde, ``{On the Origin of Gravity and the Laws of Newton},'' {\em
  JHEP} {\bf 1104} (2011) 029,
\href{http://www.arXiv.org/abs/1001.0785}{{\tt 1001.0785}}.

\bibitem{Policastro:2001yc}
G.~Policastro, D.~Son, and A.~Starinets, ``{The Shear viscosity of strongly
  coupled N=4 supersymmetric Yang-Mills plasma},'' {\em Phys.Rev.Lett.} {\bf
  87} (2001) 081601,
\href{http://www.arXiv.org/abs/hep-th/0104066}{{\tt hep-th/0104066}}.

\bibitem{Kovtun:2004de}
P.~Kovtun, D.~T. Son, and A.~O. Starinets, ``Viscosity in strongly interacting
  quantum field theories from black hole physics,'' {\em Phys. Rev. Lett.} {\bf
  94} (2005) 111601,
\href{http://www.arXiv.org/abs/arXiv:hep-th/0405231}{{\tt
  arXiv:hep-th/0405231}}.

\bibitem{Romatschke:2007mq}
P.~Romatschke and U.~Romatschke, ``{Viscosity Information from Relativistic
  Nuclear Collisions: How Perfect is the Fluid Observed at RHIC?},'' {\em
  Phys.Rev.Lett.} {\bf 99} (2007) 172301,
\href{http://www.arXiv.org/abs/0706.1522}{{\tt 0706.1522}}.

\bibitem{Teaney:2009qa}
D.~A. Teaney, ``{Viscous Hydrodynamics and the Quark Gluon Plasma},''
\href{http://www.arXiv.org/abs/0905.2433}{{\tt 0905.2433}}.

\bibitem{CasalderreySolana:2011us}
J.~Casalderrey-Solana, H.~Liu, D.~Mateos, K.~Rajagopal, and U.~A. Wiedemann,
  ``{Gauge/String Duality, Hot QCD and Heavy Ion Collisions},''
\href{http://www.arXiv.org/abs/1101.0618}{{\tt 1101.0618}}.

\bibitem{Shuryak:2011aa}
E.~Shuryak, ``{Toward the AdS/CFT Dual of the 'Little Bang'},'' {\em J.Phys.}
  {\bf G39} (2012) 054001,
\href{http://www.arXiv.org/abs/1112.2573}{{\tt 1112.2573}}.

\bibitem{DeWolfe:2013cua}
O.~DeWolfe, S.~S. Gubser, C.~Rosen, and D.~Teaney, ``{Heavy ions and string
  theory},''
\href{http://www.arXiv.org/abs/1304.7794}{{\tt 1304.7794}}.

\bibitem{PhD:Damour}
T.~Damour, {\em Quelques propri\'et\'es, m\'echaniques, \'electromagn\'eiques,
  thermodynamiques et quantiques des trous noirs}.
\newblock PhD thesis, Universit\'e Pierre et Marie Curie, Paris, 1979.

\bibitem{Son:2008ye}
D.~T. Son, ``{Toward an AdS/cold atoms correspondence: a geometric realization
  of the Schroedinger symmetry},'' {\em Phys. Rev.} {\bf D78} (2008) 046003,
\href{http://www.arXiv.org/abs/0804.3972}{{\tt 0804.3972}}.

\bibitem{Balasubramanian:2008dm}
K.~Balasubramanian and J.~McGreevy, ``{Gravity duals for non-relativistic
  CFTs},'' {\em Phys. Rev. Lett.} {\bf 101} (2008) 061601,
\href{http://www.arXiv.org/abs/0804.4053}{{\tt 0804.4053}}.

\bibitem{Adams:2008wt}
A.~Adams, K.~Balasubramanian, and J.~McGreevy, ``{Hot Spacetimes for Cold
  Atoms},'' {\em JHEP} {\bf 11} (2008) 059,
\href{http://www.arXiv.org/abs/0807.1111}{{\tt 0807.1111}}.

\bibitem{Kachru:2008yh}
S.~Kachru, X.~Liu, and M.~Mulligan, ``{Gravity Duals of Lifshitz-like Fixed
  Points},'' {\em Phys. Rev.} {\bf D78} (2008) 106005,
\href{http://www.arXiv.org/abs/0808.1725}{{\tt 0808.1725}}.

\bibitem{Gubser:2008px}
S.~S. Gubser, ``{Breaking an Abelian gauge symmetry near a black hole
  horizon},'' {\em Phys.Rev.} {\bf D78} (2008) 065034,
\href{http://www.arXiv.org/abs/0801.2977}{{\tt 0801.2977}}.

\bibitem{Hartnoll:2008vx}
S.~A. Hartnoll, C.~P. Herzog, and G.~T. Horowitz, ``{Building a Holographic
  Superconductor},'' {\em Phys. Rev. Lett.} {\bf 101} (2008) 031601,
\href{http://www.arXiv.org/abs/0803.3295}{{\tt 0803.3295}}.

\bibitem{Hartnoll:2008kx}
S.~A. Hartnoll, C.~P. Herzog, and G.~T. Horowitz, ``{Holographic
  Superconductors},'' {\em JHEP} {\bf 0812} (2008) 015,
\href{http://www.arXiv.org/abs/0810.1563}{{\tt 0810.1563}}.

\bibitem{Hartnoll:2009ns}
S.~A. Hartnoll, J.~Polchinski, E.~Silverstein, and D.~Tong, ``{Towards strange
  metallic holography},'' {\em JHEP} {\bf 1004} (2010) 120,
\href{http://www.arXiv.org/abs/0912.1061}{{\tt 0912.1061}}.

\bibitem{Liu:2009dm}
H.~Liu, J.~McGreevy, and D.~Vegh, ``{Non-Fermi liquids from holography},'' {\em
  Phys.Rev.} {\bf D83} (2011) 065029,
\href{http://www.arXiv.org/abs/0903.2477}{{\tt 0903.2477}}.

\bibitem{Faulkner:2010zz}
T.~Faulkner, N.~Iqbal, H.~Liu, J.~McGreevy, and D.~Vegh, ``{Strange metal
  transport realized by gauge/gravity duality},'' {\em Science} {\bf 329}
  (2010)
1043--1047.

\bibitem{Faulkner:2011tm}
T.~Faulkner, N.~Iqbal, H.~Liu, J.~McGreevy, and D.~Vegh, ``{Holographic
  non-Fermi liquid fixed points},'' {\em Phil. Trans. Roy. Soc.} {\bf A 369}
  (2011) 1640,
\href{http://www.arXiv.org/abs/1101.0597}{{\tt 1101.0597}}.

\bibitem{Iqbal:2008by}
N.~Iqbal and H.~Liu, ``{Universality of the hydrodynamic limit in AdS/CFT and
  the membrane paradigm},'' {\em Phys.Rev.} {\bf D79} (2009) 025023,
\href{http://www.arXiv.org/abs/0809.3808}{{\tt 0809.3808}}.

\bibitem{Bredberg:2010ky}
I.~Bredberg, C.~Keeler, V.~Lysov, and A.~Strominger, ``{Wilsonian Approach to
  Fluid/Gravity Duality},'' {\em JHEP} {\bf 1103} (2011) 141,
\href{http://www.arXiv.org/abs/1006.1902}{{\tt 1006.1902}}.

\bibitem{Compere:2011dx}
G.~Compere, P.~McFadden, K.~Skenderis, and M.~Taylor, ``{The Holographic fluid
  dual to vacuum Einstein gravity},'' {\em JHEP} {\bf 1107} (2011) 050,
\href{http://www.arXiv.org/abs/1103.3022}{{\tt 1103.3022}}.

\bibitem{Bredberg:2011jq}
I.~Bredberg, C.~Keeler, V.~Lysov, and A.~Strominger, ``{From Navier-Stokes To
  Einstein},'' {\em JHEP} {\bf 1207} (2012) 146,
\href{http://www.arXiv.org/abs/1101.2451}{{\tt 1101.2451}}.

\bibitem{Compere:2012mt}
G.~Compere, P.~McFadden, K.~Skenderis, and M.~Taylor, ``{The relativistic fluid
  dual to vacuum Einstein gravity},'' {\em JHEP} {\bf 1203} (2012) 076,
\href{http://www.arXiv.org/abs/1201.2678}{{\tt 1201.2678}}.

\bibitem{Thorne:1986iy}
K.~S. Thorne, R.~Price, and D.~Macdonald,
``Black holes: The membrane paradigm,''.

\bibitem{McGreevy:2009xe}
J.~McGreevy, ``{Holographic duality with a view toward many-body physics},''
  {\em Adv.High Energy Phys.} {\bf 2010} (2010) 723105,
\href{http://www.arXiv.org/abs/0909.0518}{{\tt 0909.0518}}.

\bibitem{Sachdev:2010uj}
S.~Sachdev, ``{Strange metals and the AdS/CFT correspondence},'' {\em
  J.Stat.Mech.} {\bf 1011} (2010) P11022,
\href{http://www.arXiv.org/abs/1010.0682}{{\tt 1010.0682}}.

\bibitem{Hartnoll:2011fn}
S.~A. Hartnoll, ``{Horizons, holography and condensed matter},''
\href{http://www.arXiv.org/abs/1106.4324}{{\tt 1106.4324}}.

\bibitem{Sachdev:2011wg}
S.~Sachdev, ``{What can gauge-gravity duality teach us about condensed matter
  physics?},'' {\em Ann.Rev.Condensed Matter Phys.} {\bf 3} (2012) 9--33,
\href{http://www.arXiv.org/abs/1108.1197}{{\tt 1108.1197}}.

\bibitem{Iqbal:2011ae}
N.~Iqbal, H.~Liu, and M.~Mezei, ``{Lectures on holographic non-Fermi liquids
  and quantum phase transitions},''
\href{http://www.arXiv.org/abs/1110.3814}{{\tt 1110.3814}}.

\bibitem{Son:2012zz}
D.~T. Son, ``{Holography for strongly coupled media},'' {\em Lect.Notes Phys.}
  {\bf 851} (2012)
147--163.

\bibitem{Grumiller:2008qz}
D.~Grumiller and N.~Johansson, ``{Instability in cosmological topologically
  massive gravity at the chiral point},'' {\em JHEP} {\bf 07} (2008) 134,
\href{http://www.arXiv.org/abs/0805.2610}{{\tt 0805.2610}}.

\bibitem{Flohr:2001zs}
M.~Flohr, ``{Bits and pieces in logarithmic conformal field theory},'' {\em
  Int. J. Mod. Phys.} {\bf A18} (2003) 4497--4592,
\href{http://www.arXiv.org/abs/hep-th/0111228}{{\tt hep-th/0111228}}.

\bibitem{Gaberdiel:2001tr}
M.~R. Gaberdiel, ``{An algebraic approach to logarithmic conformal field
  theory},'' {\em Int. J. Mod. Phys.} {\bf A18} (2003) 4593--4638,
\href{http://www.arXiv.org/abs/hep-th/0111260}{{\tt hep-th/0111260}}.

\bibitem{lcft}
J.~Cardy and et~al., ``Logarithmic conformal field theories,'' {\em J. Phys.}
  {\bf A46} (2013). special issue {\em Logarithmic conformal field theories}
  edited by A. Gainutdinov, D. Ridout and I. Runkel.

\bibitem{Grumiller:2013at}
D.~Grumiller, W.~Riedler, J.~Rosseel, and T.~Zojer, ``{Holographic applications
  of logarithmic conformal field theories},'' {\em J. Phys. A: Math. Theor.}
  {\bf 46} (2013) 494002,
\href{http://www.arXiv.org/abs/1302.0280}{{\tt 1302.0280}}.

\bibitem{Susskind:1998vk}
L.~Susskind, ``{Holography in the flat space limit},''
\href{http://www.arXiv.org/abs/hep-th/9901079}{{\tt hep-th/9901079}}.

\bibitem{Polchinski:1999ry}
J.~Polchinski, ``{S matrices from AdS space-time},''
\href{http://www.arXiv.org/abs/hep-th/9901076}{{\tt hep-th/9901076}}.

\bibitem{Giddings:1999jq}
S.~B. Giddings, ``{Flat space scattering and bulk locality in the AdS / CFT
  correspondence},'' {\em Phys.Rev.} {\bf D61} (2000) 106008,
\href{http://www.arXiv.org/abs/hep-th/9907129}{{\tt hep-th/9907129}}.

\bibitem{Gary:2009ae}
M.~Gary, S.~B. Giddings, and J.~Penedones, ``{Local bulk S-matrix elements and
  CFT singularities},'' {\em Phys.Rev.} {\bf D80} (2009) 085005,
\href{http://www.arXiv.org/abs/0903.4437}{{\tt 0903.4437}}.

\bibitem{Gary:2009mi}
M.~Gary and S.~B. Giddings, ``{The Flat space S-matrix from the AdS/CFT
  correspondence?},'' {\em Phys.Rev.} {\bf D80} (2009) 046008,
\href{http://www.arXiv.org/abs/0904.3544}{{\tt 0904.3544}}.

\bibitem{Barnich:2006av}
G.~Barnich and G.~Compere, ``{Classical central extension for asymptotic
  symmetries at null infinity in three spacetime dimensions},'' {\em
  Class.Quant.Grav.} {\bf 24} (2007) F15--F23,
\href{http://www.arXiv.org/abs/gr-qc/0610130}{{\tt gr-qc/0610130}}.

\bibitem{Bondi:1962}
H.~Bondi, M.~van~der Burg, and A.~Metzner, ``Gravitational waves in general
  relativity {VII.} {W}aves from axi-symmetric isolated systems,'' {\em Proc.
  Roy. Soc. London} {\bf A269} (1962) 21--51.

\bibitem{Sachs:1962}
R.~Sachs, ``Asymptotic symmetries in gravitational theory,'' {\em Phys. Rev.}
  {\bf 128} (1962) 2851--2864.

\bibitem{Bagchi:2009my}
A.~Bagchi and R.~Gopakumar, ``{Galilean Conformal Algebras and AdS/CFT},'' {\em
  JHEP} {\bf 0907} (2009) 037,
\href{http://www.arXiv.org/abs/0902.1385}{{\tt 0902.1385}}.

\bibitem{Bagchi:2009pe}
A.~Bagchi, R.~Gopakumar, I.~Mandal, and A.~Miwa, ``{GCA in 2d},'' {\em JHEP}
  {\bf 1008} (2010) 004,
\href{http://www.arXiv.org/abs/0912.1090}{{\tt 0912.1090}}.

\bibitem{Bagchi:2010zz}
A.~Bagchi, ``{Correspondence between Asymptotically Flat Spacetimes and
  Nonrelativistic Conformal Field Theories},'' {\em Phys.Rev.Lett.} {\bf 105}
  (2010)
171601.

\bibitem{Bagchi:2012yk}
A.~Bagchi, S.~Detournay, and D.~Grumiller, ``{Flat-Space Chiral Gravity},''
  {\em Phys.Rev.Lett.} {\bf 109} (2012) 151301,
\href{http://www.arXiv.org/abs/1208.1658}{{\tt 1208.1658}}.

\bibitem{Witten:2007kt}
E.~Witten, ``{Three-Dimensional Gravity Revisited},''
\href{http://www.arXiv.org/abs/0706.3359}{{\tt 0706.3359}}.

\bibitem{Li:2008dq}
W.~Li, W.~Song, and A.~Strominger, ``{Chiral Gravity in Three Dimensions},''
  {\em JHEP} {\bf 04} (2008) 082,
\href{http://www.arXiv.org/abs/0801.4566}{{\tt 0801.4566}}.

\bibitem{Maloney:2009ck}
A.~Maloney, W.~Song, and A.~Strominger, ``{Chiral Gravity, Log Gravity and
  Extremal CFT},'' {\em Phys. Rev.} {\bf D81} (2010) 064007,
\href{http://www.arXiv.org/abs/0903.4573}{{\tt 0903.4573}}.

\bibitem{Cornalba:2002fi}
L.~Cornalba and M.~S. Costa, ``{A New cosmological scenario in string
  theory},'' {\em Phys.Rev.} {\bf D66} (2002) 066001,
\href{http://www.arXiv.org/abs/hep-th/0203031}{{\tt hep-th/0203031}}.

\bibitem{Cornalba:2003kd}
L.~Cornalba and M.~S. Costa, ``{Time dependent orbifolds and string
  cosmology},'' {\em Fortsch.Phys.} {\bf 52} (2004) 145--199,
\href{http://www.arXiv.org/abs/hep-th/0310099}{{\tt hep-th/0310099}}.

\bibitem{Barnich:2012xq}
G.~Barnich, ``{Entropy of three-dimensional asymptotically flat cosmological
  solutions},'' {\em JHEP} {\bf 1210} (2012) 095,
\href{http://www.arXiv.org/abs/1208.4371}{{\tt 1208.4371}}.

\bibitem{Bagchi:2012xr}
A.~Bagchi, S.~Detournay, R.~Fareghbal, and J.~Simon, ``{Holography of 3d Flat
  Cosmological Horizons},'' {\em Phys. Rev. Lett.} {\bf 110} (2013) 141302,
\href{http://www.arXiv.org/abs/1208.4372}{{\tt 1208.4372}}.

\bibitem{Bagchi:2013lma}
A.~Bagchi, S.~Detournay, D.~Grumiller, and J.~Simon, ``{Cosmic evolution from
  phase transition of 3-dimensional flat space},'' {\em Phys.Rev.Lett.} {\bf
  111} (2013) 181301,
\href{http://www.arXiv.org/abs/1305.2919}{{\tt 1305.2919}}.

\bibitem{Barnich:2010eb}
G.~Barnich and C.~Troessaert, ``{Aspects of the BMS/CFT correspondence},'' {\em
  JHEP} {\bf 1005} (2010) 062,
\href{http://www.arXiv.org/abs/1001.1541}{{\tt 1001.1541}}.

\bibitem{Barnich:2011mi}
G.~Barnich and C.~Troessaert, ``{BMS charge algebra},'' {\em JHEP} {\bf 1112}
  (2011) 105,
\href{http://www.arXiv.org/abs/1106.0213}{{\tt 1106.0213}}.

\bibitem{Barnich:2012aw}
G.~Barnich, A.~Gomberoff, and H.~A. Gonzalez, ``{The Flat limit of three
  dimensional asymptotically anti-de Sitter spacetimes},'' {\em Phys.Rev.} {\bf
  D86} (2012) 024020,
\href{http://www.arXiv.org/abs/1204.3288}{{\tt 1204.3288}}.

\bibitem{Barnich:2012rz}
G.~Barnich, A.~Gomberoff, and H.~A. Gonzalez, ``{BMS$_3$ invariant two
  dimensional field theories as flat limit of Liouville},'' {\em Phys. Rev.}
  {\bf D87:124032,} (2013)
\href{http://www.arXiv.org/abs/1210.0731}{{\tt 1210.0731}}.

\bibitem{Bagchi:2012cy}
A.~Bagchi and R.~Fareghbal, ``{BMS/GCA Redux: Towards Flatspace Holography from
  Non-Relativistic Symmetries},'' {\em JHEP} {\bf 1210} (2012) 092,
\href{http://www.arXiv.org/abs/1203.5795}{{\tt 1203.5795}}.

\bibitem{Barnich:2013yka}
G.~Barnich and H.~A. Gonzalez, ``{Dual dynamics of three dimensional
  asymptotically flat Einstein gravity at null infinity},'' {\em JHEP} {\bf
  1305} (2013) 016,
\href{http://www.arXiv.org/abs/1303.1075}{{\tt 1303.1075}}.

\bibitem{Barnich:2013axa}
G.~Barnich and C.~Troessaert, ``{Comments on holographic current algebras and
  asymptotically flat four dimensional spacetimes at null infinity},'' {\em
  JHEP} {\bf 1311} (2013) 003,
\href{http://www.arXiv.org/abs/1309.0794}{{\tt 1309.0794}}.

\bibitem{Bagchi:2013qva}
A.~Bagchi and R.~Basu, ``{3D Flat Holography: Entropy and Logarithmic
  Corrections},''
\href{http://www.arXiv.org/abs/1312.5748}{{\tt 1312.5748}}.

\bibitem{Costa:2013vza}
R.~N.~C. Costa, ``{Aspects of the zero $\Lambda$ limit in the AdS/CFT
  correspondence},''
\href{http://www.arXiv.org/abs/1311.7339}{{\tt 1311.7339}}.

\bibitem{Fradkin:1986qy}
E.~Fradkin and M.~A. Vasiliev, ``{Cubic Interaction in Extended Theories of
  Massless Higher Spin Fields},'' {\em Nucl.Phys.} {\bf B291} (1987) 141.

\bibitem{Fradkin:1987ks}
E.~Fradkin and M.~A. Vasiliev, ``{On the Gravitational Interaction of Massless
  Higher Spin Fields},'' {\em Phys.Lett.} {\bf B189} (1987) 89--95.

\bibitem{Vasiliev:1990en}
M.~A. Vasiliev, ``{Consistent equation for interacting gauge fields of all
  spins in (3+1)-dimensions},'' {\em Phys.Lett.} {\bf B243} (1990) 378--382.

\bibitem{Vasiliev:2003ev}
M.~Vasiliev, ``{Nonlinear equations for symmetric massless higher spin fields
  in (A)dS(d)},'' {\em Phys.Lett.} {\bf B567} (2003) 139--151,
  \href{http://www.arXiv.org/abs/hep-th/0304049}{{\tt hep-th/0304049}}.

\bibitem{Giombi:2009wh}
S.~Giombi and X.~Yin, ``{Higher Spin Gauge Theory and Holography: The
  Three-Point Functions},'' {\em JHEP} {\bf 1009} (2010) 115,
  \href{http://www.arXiv.org/abs/0912.3462}{{\tt 0912.3462}}.

\bibitem{Giombi:2010vg}
S.~Giombi and X.~Yin, ``{Higher Spins in AdS and Twistorial Holography},'' {\em
  JHEP} {\bf 1104} (2011) 086, \href{http://www.arXiv.org/abs/1004.3736}{{\tt
  1004.3736}}.

\bibitem{Koch:2010cy}
R.~d.~M. Koch, A.~Jevicki, K.~Jin, and J.~P. Rodrigues, ``{$AdS_4/CFT_3$
  Construction from Collective Fields},'' {\em Phys.Rev.} {\bf D83} (2011)
  025006, \href{http://www.arXiv.org/abs/1008.0633}{{\tt 1008.0633}}.

\bibitem{Giombi:2011ya}
S.~Giombi and X.~Yin, ``{On Higher Spin Gauge Theory and the Critical O(N)
  Model},'' {\em Phys.Rev.} {\bf D85} (2012) 086005,
\href{http://www.arXiv.org/abs/1105.4011}{{\tt 1105.4011}}.

\bibitem{Gaberdiel:2010pz}
M.~R. Gaberdiel and R.~Gopakumar, ``{An AdS$_3$ Dual for Minimal Model CFTs},''
  {\em Phys.Rev.} {\bf D83} (2011) 066007,
  \href{http://www.arXiv.org/abs/1011.2986}{{\tt 1011.2986}}.

\bibitem{Gaberdiel:2011zw}
M.~R. Gaberdiel, R.~Gopakumar, T.~Hartman, and S.~Raju, ``{Partition Functions
  of Holographic Minimal Models},'' {\em JHEP} {\bf 1108} (2011) 077,
  \href{http://www.arXiv.org/abs/1106.1897}{{\tt 1106.1897}}.

\bibitem{Gaberdiel:2012uj}
M.~R. Gaberdiel and R.~Gopakumar, ``{Minimal Model Holography},'' {\em J.Phys.}
  {\bf A46} (2013) 214002,
\href{http://www.arXiv.org/abs/1207.6697}{{\tt 1207.6697}}.

\bibitem{Henneaux:2010xg}
M.~Henneaux and S.-J. Rey, ``{Nonlinear $W_{infinity}$ as Asymptotic Symmetry
  of Three-Dimensional Higher Spin Anti-de Sitter Gravity},'' {\em JHEP} {\bf
  1012} (2010) 007, \href{http://www.arXiv.org/abs/1008.4579}{{\tt 1008.4579}}.

\bibitem{Campoleoni:2010zq}
A.~Campoleoni, S.~Fredenhagen, S.~Pfenninger, and S.~Theisen, ``{Asymptotic
  symmetries of three-dimensional gravity coupled to higher-spin fields},''
  {\em JHEP} {\bf 1011} (2010) 007,
  \href{http://www.arXiv.org/abs/1008.4744}{{\tt 1008.4744}}.

\bibitem{Fotopoulos:2008ka}
A.~Fotopoulos and M.~Tsulaia, ``{Gauge Invariant Lagrangians for Free and
  Interacting Higher Spin Fields. A Review of the BRST formulation},'' {\em
  Int.J.Mod.Phys.} {\bf A24} (2009) 1--60,
\href{http://www.arXiv.org/abs/0805.1346}{{\tt 0805.1346}}.

\bibitem{Sagnotti:2010at}
A.~Sagnotti and M.~Taronna, ``{String Lessons for Higher-Spin Interactions},''
  {\em Nucl.Phys.} {\bf B842} (2011) 299--361,
\href{http://www.arXiv.org/abs/1006.5242}{{\tt 1006.5242}}.

\bibitem{Bekaert:2010hw}
X.~Bekaert, N.~Boulanger, and P.~Sundell, ``{How higher-spin gravity surpasses
  the spin two barrier: no-go theorems versus yes-go examples},'' {\em
  Rev.Mod.Phys.} {\bf 84} (2012) 987--1009,
\href{http://www.arXiv.org/abs/1007.0435}{{\tt 1007.0435}}.

\bibitem{Campoleoni:2011hg}
A.~Campoleoni, S.~Fredenhagen, and S.~Pfenninger, ``{Asymptotic W-symmetries in
  three-dimensional higher-spin gauge theories},'' {\em JHEP} {\bf 1109} (2011)
  113,
\href{http://www.arXiv.org/abs/1107.0290}{{\tt 1107.0290}}.

\bibitem{Ammon:2011nk}
M.~Ammon, M.~Gutperle, P.~Kraus, and E.~Perlmutter, ``{Spacetime Geometry in
  Higher Spin Gravity},'' \href{http://www.arXiv.org/abs/1106.4788}{{\tt
  1106.4788}}.

\bibitem{Anninos:2011ui}
D.~Anninos, T.~Hartman, and A.~Strominger, ``{Higher Spin Realization of the
  dS/CFT Correspondence},''
\href{http://www.arXiv.org/abs/1108.5735}{{\tt 1108.5735}}.

\bibitem{Maldacena:2011jn}
J.~Maldacena and A.~Zhiboedov, ``{Constraining Conformal Field Theories with A
  Higher Spin Symmetry},'' {\em J.Phys.} {\bf A46} (2013) 214011,
\href{http://www.arXiv.org/abs/1112.1016}{{\tt 1112.1016}}.

\bibitem{Ammon:2012wc}
M.~Ammon, M.~Gutperle, P.~Kraus, and E.~Perlmutter, ``{Black holes in three
  dimensional higher spin gravity: A review},'' {\em J.Phys.} {\bf A46} (2013)
  214001,
\href{http://www.arXiv.org/abs/1208.5182}{{\tt 1208.5182}}.

\bibitem{Vasiliev:2012vf}
M.~A. Vasiliev, ``{Holography, Unfolding and Higher-Spin Theory},'' {\em
  J.Phys.} {\bf A46} (2013) 214013,
\href{http://www.arXiv.org/abs/1203.5554}{{\tt 1203.5554}}.

\bibitem{Maldacena:2012sf}
J.~Maldacena and A.~Zhiboedov, ``{Constraining conformal field theories with a
  slightly broken higher spin symmetry},'' {\em Class.Quant.Grav.} {\bf 30}
  (2013) 104003,
\href{http://www.arXiv.org/abs/1204.3882}{{\tt 1204.3882}}.

\bibitem{Afshar:2013vka}
H.~Afshar, A.~Bagchi, R.~Fareghbal, D.~Grumiller, and J.~Rosseel, ``{Higher
  spin theory in 3-dimensional flat space},'' {\em Phys.Rev.Lett.} {\bf 111}
  (2013) 121603,
\href{http://www.arXiv.org/abs/1307.4768}{{\tt 1307.4768}}.

\bibitem{Gonzalez:2013oaa}
H.~A. Gonzalez, J.~Matulich, M.~Pino, and R.~Troncoso, ``{Asymptotically flat
  spacetimes in three-dimensional higher spin gravity},'' {\em JHEP} {\bf 1309}
  (2013) 016,
\href{http://www.arXiv.org/abs/1307.5651}{{\tt 1307.5651}}.

\bibitem{Gary:2012ms}
M.~Gary, D.~Grumiller, and R.~Rashkov, ``{Towards non-AdS holography in
  3-dimensional higher spin gravity},'' {\em JHEP} {\bf 1203} (2012) 022,
\href{http://www.arXiv.org/abs/1201.0013}{{\tt 1201.0013}}.

\bibitem{Nielsen:2000}
N.~Nielsen and I.~L. Chuang, {\em Quantum Computation and Quantum Information}.
\newblock Cambridge University Press, 2000.

\bibitem{Sorkin:1985bu}
R.~D. Sorkin, ``{ON THE ENTROPY OF THE VACUUM OUTSIDE A HORIZON},''
\href{http://www.arXiv.org/abs/1402.3589}{{\tt 1402.3589}}.

\bibitem{Bombelli:1986rw}
L.~Bombelli, R.~K. Koul, J.~Lee, and R.~D. Sorkin, ``{A Quantum Source of
  Entropy for Black Holes},'' {\em Phys.Rev.} {\bf D34} (1986)
373--383.

\bibitem{Srednicki:1993im}
M.~Srednicki, ``{Entropy and area},'' {\em Phys.Rev.Lett.} {\bf 71} (1993)
  666--669,
\href{http://www.arXiv.org/abs/hep-th/9303048}{{\tt hep-th/9303048}}.

\bibitem{Solodukhin:2011gn}
S.~N. Solodukhin, ``{Entanglement entropy of black holes},'' {\em Living
  Rev.Rel.} {\bf 14} (2011) 8,
\href{http://www.arXiv.org/abs/1104.3712}{{\tt 1104.3712}}.

\bibitem{Vidal:2002rm}
G.~Vidal, J.~Latorre, E.~Rico, and A.~Kitaev, ``{Entanglement in quantum
  critical phenomena},'' {\em Phys.Rev.Lett.} {\bf 90} (2003) 227902,
\href{http://www.arXiv.org/abs/quant-ph/0211074}{{\tt quant-ph/0211074}}.

\bibitem{Calabrese:2004eu}
P.~Calabrese and J.~L. Cardy, ``{Entanglement entropy and quantum field
  theory},'' {\em J.Stat.Mech.} {\bf 0406} (2004) P06002,
\href{http://www.arXiv.org/abs/hep-th/0405152}{{\tt hep-th/0405152}}.

\bibitem{Kitaev:2005dm}
A.~Kitaev and J.~Preskill, ``{Topological entanglement entropy},'' {\em
  Phys.Rev.Lett.} {\bf 96} (2006) 110404,
\href{http://www.arXiv.org/abs/hep-th/0510092}{{\tt hep-th/0510092}}.

\bibitem{Levin:2006zz}
M.~Levin and X.-G. Wen, ``{Detecting Topological Order in a Ground State Wave
  Function},'' {\em Phys.Rev.Lett.} {\bf 96} (2006)
110405.

\bibitem{Ryu:2006bv}
S.~Ryu and T.~Takayanagi, ``{Holographic derivation of entanglement entropy
  from AdS/CFT},'' {\em Phys.Rev.Lett.} {\bf 96} (2006) 181602,
\href{http://www.arXiv.org/abs/hep-th/0603001}{{\tt hep-th/0603001}}.

\bibitem{Ryu:2006ef}
S.~Ryu and T.~Takayanagi, ``{Aspects of Holographic Entanglement Entropy},''
  {\em JHEP} {\bf 0608} (2006) 045,
\href{http://www.arXiv.org/abs/hep-th/0605073}{{\tt hep-th/0605073}}.

\bibitem{Nishioka:2009un}
T.~Nishioka, S.~Ryu, and T.~Takayanagi, ``{Holographic Entanglement Entropy: An
  Overview},'' {\em J.Phys.} {\bf A42} (2009) 504008,
\href{http://www.arXiv.org/abs/0905.0932}{{\tt 0905.0932}}.

\bibitem{Hayden:2007cs}
P.~Hayden and J.~Preskill, ``{Black holes as mirrors: Quantum information in
  random subsystems},'' {\em JHEP} {\bf 0709} (2007) 120,
\href{http://www.arXiv.org/abs/0708.4025}{{\tt 0708.4025}}.

\bibitem{Harlow:2013tf}
D.~Harlow and P.~Hayden, ``{Quantum Computation vs. Firewalls},'' {\em JHEP}
  {\bf 1306} (2013) 085,
\href{http://www.arXiv.org/abs/1301.4504}{{\tt 1301.4504}}.

\bibitem{Aman:2003ug}
J.~E. Aman, I.~Bengtsson, and N.~Pidokrajt, ``{Geometry of black hole
  thermodynamics},'' {\em Gen.Rel.Grav.} {\bf 35} (2003) 1733,
\href{http://www.arXiv.org/abs/gr-qc/0304015}{{\tt gr-qc/0304015}}.

\bibitem{Arcioni:2004ww}
G.~Arcioni and E.~Lozano-Tellechea, ``Stability and critical phenomena of black
  holes and black rings,'' {\em Phys. Rev.} {\bf D72} (2005) 104021,
\href{http://www.arXiv.org/abs/hep-th/0412118}{{\tt hep-th/0412118}}.

\bibitem{Aman:2005xk}
J.~E. Aman and N.~Pidokrajt, ``Geometry of higher-dimensional black hole
  thermodynamics,'' {\em Phys. Rev.} {\bf D73} (2006) 024017,
\href{http://www.arXiv.org/abs/hep-th/0510139}{{\tt hep-th/0510139}}.

\bibitem{Shen:2005nu}
J.-y. Shen, R.-G. Cai, B.~Wang, and R.-K. Su, ``Thermodynamic geometry and
  critical behavior of black holes,'' {\em Int. J. Mod. Phys.} {\bf A22} (2007)
  11--27,
\href{http://www.arXiv.org/abs/gr-qc/0512035}{{\tt gr-qc/0512035}}.

\bibitem{Sarkar:2006tg}
T.~Sarkar, G.~Sengupta, and B.~Nath~Tiwari, ``{On the thermodynamic geometry of
  BTZ black holes},'' {\em JHEP} {\bf 0611} (2006) 015,
\href{http://www.arXiv.org/abs/hep-th/0606084}{{\tt hep-th/0606084}}.

\bibitem{Alvarez:2008wa}
J.~L. Alvarez, H.~Quevedo, and A.~Sanchez, ``{Unified geometric description of
  black hole thermodynamics},'' {\em Phys.Rev.} {\bf D77} (2008) 084004,
\href{http://www.arXiv.org/abs/0801.2279}{{\tt 0801.2279}}.

\bibitem{Ruppeiner:2008kd}
G.~Ruppeiner, ``{Thermodynamic curvature and phase transitions in Kerr-Newman
  black holes},'' {\em Phys.Rev.} {\bf D78} (2008) 024016,
\href{http://www.arXiv.org/abs/0802.1326}{{\tt 0802.1326}}.

\bibitem{Ruppeiner:1979}
G.~Ruppeiner, ``{Thermodynamics: A Riemannian geometric model},'' {\em Phys.
  Rev.} {\bf A20} (1979) 1608.

\bibitem{RevModPhys.67.605}
G.~Ruppeiner, ``Riemannian geometry in thermodynamic fluctuation theory,'' {\em
  Rev. Mod. Phys.} {\bf 67} (Jul, 1995) 605--659.

\bibitem{RevModPhys.68.313}
G.~Ruppeiner, ``Erratum: Riemannian geometry in thermodynamic fluctuation
  theory,'' {\em Rev. Mod. Phys.} {\bf 68} (Jan, 1996) 313.

\bibitem{Henneaux:1984ji}
M.~Henneaux and C.~Teitelboim, ``{The Cosmological Constant as a Canonical
  Variable},'' {\em Phys.Lett.} {\bf B143} (1984)
415--420.

\bibitem{Henneaux:1985tv}
M.~Henneaux and C.~Teitelboim, ``{Asymptotically anti-De Sitter Spaces},'' {\em
  Commun.Math.Phys.} {\bf 98} (1985)
391--424.

\bibitem{Brown:1987dd}
J.~D. Brown and C.~Teitelboim, ``{Dynamical Neutralization of the Cosmological
  Constant},'' {\em Phys.Lett.} {\bf B195} (1987)
177--182.

\bibitem{Brown:1988kg}
J.~D. Brown and C.~Teitelboim, ``{Neutralization of the Cosmological Constant
  by Membrane Creation},'' {\em Nucl.Phys.} {\bf B297} (1988)
787--836.

\bibitem{Bousso:2000xa}
R.~Bousso and J.~Polchinski, ``{Quantization of four form fluxes and dynamical
  neutralization of the cosmological constant},'' {\em JHEP} {\bf 0006} (2000)
  006,
\href{http://www.arXiv.org/abs/hep-th/0004134}{{\tt hep-th/0004134}}.

\bibitem{Gomberoff:2003zh}
A.~Gomberoff, M.~Henneaux, C.~Teitelboim, and F.~Wilczek, ``{Thermal decay of
  the cosmological constant into black holes},'' {\em Phys.Rev.} {\bf D69}
  (2004) 083520,
\href{http://www.arXiv.org/abs/hep-th/0311011}{{\tt hep-th/0311011}}.

\bibitem{Caldarelli:1999xj}
M.~M. Caldarelli, G.~Cognola, and D.~Klemm, ``{Thermodynamics of
  Kerr-Newman-AdS black holes and conformal field theories},'' {\em
  Class.Quant.Grav.} {\bf 17} (2000) 399--420,
\href{http://www.arXiv.org/abs/hep-th/9908022}{{\tt hep-th/9908022}}.

\bibitem{Kastor:2009wy}
D.~Kastor, S.~Ray, and J.~Traschen, ``{Enthalpy and the Mechanics of AdS Black
  Holes},'' {\em Class.Quant.Grav.} {\bf 26} (2009) 195011,
\href{http://www.arXiv.org/abs/0904.2765}{{\tt 0904.2765}}.

\bibitem{Cvetic:2010jb}
M.~Cvetic, G.~Gibbons, D.~Kubiznak, and C.~Pope, ``{Black Hole Enthalpy and an
  Entropy Inequality for the Thermodynamic Volume},'' {\em Phys.Rev.} {\bf D84}
  (2011) 024037,
\href{http://www.arXiv.org/abs/1012.2888}{{\tt 1012.2888}}.

\bibitem{Dolan:2010ha}
B.~P. Dolan, ``{The Cosmological Constant and black hole Thermodynamic
  Potentials},'' {\em Class.Quant.Grav.} {\bf 28} (2011) 125020,
\href{http://www.arXiv.org/abs/1008.5023}{{\tt 1008.5023}}.

\bibitem{Dolan:2011xt}
B.~P. Dolan, ``{Pressure and volume in the first law of black hole
  thermodynamics},'' {\em Class.Quant.Grav.} {\bf 28} (2011) 235017,
\href{http://www.arXiv.org/abs/1106.6260}{{\tt 1106.6260}}.

\bibitem{Kubiznak:2012wp}
D.~Kubiznak and R.~B. Mann, ``{P-V criticality of charged AdS black holes},''
  {\em JHEP} {\bf 1207} (2012) 033,
\href{http://www.arXiv.org/abs/1205.0559}{{\tt 1205.0559}}.

\bibitem{Dolan:2013dga}
B.~P. Dolan, ``{The compressibility of rotating black holes in
  $D$-dimensions},'' {\em Class.Quant.Grav.} {\bf 31} (2014) 035022,
\href{http://www.arXiv.org/abs/1308.5403}{{\tt 1308.5403}}.

\bibitem{Dolan:2013ft}
B.~P. Dolan, D.~Kastor, D.~Kubiznak, R.~B. Mann, and J.~Traschen,
  ``{Thermodynamic Volumes and Isoperimetric Inequalities for de Sitter Black
  Holes},'' {\em Phys.Rev.} {\bf D87} (2013) 104017,
\href{http://www.arXiv.org/abs/1301.5926}{{\tt 1301.5926}}.

\bibitem{Altamirano:2014tva}
N.~Altamirano, D.~Kubiznak, R.~B. Mann, and Z.~Sherkatghanad, ``{Thermodynamics
  of rotating black holes and black rings: phase transitions and thermodynamic
  volume},''
\href{http://www.arXiv.org/abs/1401.2586}{{\tt 1401.2586}}.

\bibitem{Guica:2008mu}
M.~Guica, T.~Hartman, W.~Song, and A.~Strominger, ``{The Kerr/CFT
  Correspondence},'' {\em Phys.Rev.} {\bf D80} (2009) 124008,
\href{http://www.arXiv.org/abs/0809.4266}{{\tt 0809.4266}}.

\bibitem{Lu:2008jk}
H.~Lu, J.~Mei, and C.~Pope, ``{Kerr/CFT Correspondence in Diverse
  Dimensions},'' {\em JHEP} {\bf 0904} (2009) 054,
\href{http://www.arXiv.org/abs/0811.2225}{{\tt 0811.2225}}.

\bibitem{Azeyanagi:2008dk}
T.~Azeyanagi, N.~Ogawa, and S.~Terashima, ``{The Kerr/CFT Correspondence and
  String Theory},'' {\em Phys.Rev.} {\bf D79} (2009) 106009,
\href{http://www.arXiv.org/abs/0812.4883}{{\tt 0812.4883}}.

\bibitem{Bredberg:2009pv}
I.~Bredberg, T.~Hartman, W.~Song, and A.~Strominger, ``{Black Hole
  Superradiance From Kerr/CFT},'' {\em JHEP} {\bf 1004} (2010) 019,
\href{http://www.arXiv.org/abs/0907.3477}{{\tt 0907.3477}}.

\bibitem{Cvetic:2009jn}
M.~Cvetic and F.~Larsen, ``{Greybody Factors and Charges in Kerr/CFT},'' {\em
  JHEP} {\bf 0909} (2009) 088,
\href{http://www.arXiv.org/abs/0908.1136}{{\tt 0908.1136}}.

\bibitem{Castro:2009jf}
A.~Castro and F.~Larsen, ``{Near Extremal Kerr Entropy from AdS(2) Quantum
  Gravity},'' {\em JHEP} {\bf 0912} (2009) 037,
\href{http://www.arXiv.org/abs/0908.1121}{{\tt 0908.1121}}.

\bibitem{Dias:2009ex}
O.~J. Dias, H.~S. Reall, and J.~E. Santos, ``{Kerr-CFT and gravitational
  perturbations},'' {\em JHEP} {\bf 0908} (2009) 101,
\href{http://www.arXiv.org/abs/0906.2380}{{\tt 0906.2380}}.

\bibitem{Amsel:2009ev}
A.~J. Amsel, G.~T. Horowitz, D.~Marolf, and M.~M. Roberts, ``{No Dynamics in
  the Extremal Kerr Throat},''
\href{http://www.arXiv.org/abs/0906.2376}{{\tt 0906.2376}}.

\bibitem{Castro:2010fd}
A.~Castro, A.~Maloney, and A.~Strominger, ``{Hidden Conformal Symmetry of the
  Kerr Black Hole},'' {\em Phys.Rev.} {\bf D82} (2010) 024008,
\href{http://www.arXiv.org/abs/1004.0996}{{\tt 1004.0996}}.

\bibitem{Guica:2010ej}
M.~Guica and A.~Strominger, ``{Microscopic Realization of the Kerr/CFT
  Correspondence},'' {\em JHEP} {\bf 1102} (2011) 010,
\href{http://www.arXiv.org/abs/1009.5039}{{\tt 1009.5039}}.

\bibitem{Compere:2012jk}
G.~Compere, ``{The Kerr/CFT correspondence and its extensions: a comprehensive
  review},'' {\em Living Rev.Rel.} {\bf 15} (2012) 11,
\href{http://www.arXiv.org/abs/1203.3561}{{\tt 1203.3561}}.

\bibitem{Bardeen:1999px}
J.~M. Bardeen and G.~T. Horowitz, ``{The extreme Kerr throat geometry: A vacuum
  analog of AdS(2) x S(2)},'' {\em Phys. Rev.} {\bf D60} (1999) 104030,
\href{http://www.arXiv.org/abs/hep-th/9905099}{{\tt hep-th/9905099}}.

\bibitem{Thorne:1974ve}
K.~S. Thorne, ``{Disk accretion onto a black hole. 2. Evolution of the hole},''
  {\em Astrophys. J.} {\bf 191} (1974)
507.

\bibitem{McClintock:2006xd}
J.~E. McClintock, R.~Shafee, R.~Narayan, R.~A. Remillard, S.~W. Davis, {\em et
  al.}, ``{The Spin of the Near-Extreme Kerr Black Hole GRS 1915+105},'' {\em
  Astrophys.J.} {\bf 652} (2006) 518--539,
\href{http://www.arXiv.org/abs/astro-ph/0606076}{{\tt astro-ph/0606076}}.

\bibitem{McClintock:2009as}
J.~E. McClintock and R.~A. Remillard, ``{Measuring the Spins of Stellar-Mass
  Black Holes},''
\href{http://www.arXiv.org/abs/0902.3488}{{\tt 0902.3488}}.

\bibitem{Fender:2010tk}
R.~Fender, E.~Gallo, and D.~Russell, ``{No evidence for black hole spin
  powering of jets in X-ray binaries},'' {\em Mon.Not.Roy.Astron.Soc.} {\bf
  406} (2010) 1425--1434,
\href{http://www.arXiv.org/abs/1003.5516}{{\tt 1003.5516}}.

\bibitem{Lunin:2001jy}
O.~Lunin and S.~D. Mathur, ``{AdS / CFT duality and the black hole information
  paradox},'' {\em Nucl.Phys.} {\bf B623} (2002) 342--394,
\href{http://www.arXiv.org/abs/hep-th/0109154}{{\tt hep-th/0109154}}.

\bibitem{Lin:2004nb}
H.~Lin, O.~Lunin, and J.M.~Maldacena,``{Bubbling AdS space and 1/2 BPS geometries},"
{\em JHEP}{\bf 10} (2004) 025
\href{http://www.arxiv.org/abs/hep-th/0409174}{{\tt hep-th/0409174}}.

\bibitem{Grant:2005qc}
L.~Grant, L.~Maoz, J.~Marsano, K.~Papadodimas, and V.S.~Rychkov,
``{Minisuperspace quantization of 'Bubbling AdS' and free fermion droplets},"
{\em JHEP}{\bf 08} (2005) 025
\href{http://www.arxiv.org/abs/hep-th/0505079}{{\tt hep-th/0505079}}. 

\bibitem{Mathur:2005zp}
S.~D. Mathur, ``{The Fuzzball proposal for black holes: An Elementary
  review},'' {\em Fortsch.Phys.} {\bf 53} (2005) 793--827,
\href{http://www.arXiv.org/abs/hep-th/0502050}{{\tt hep-th/0502050}}.

\bibitem{Bena:2007kg}
I.~Bena and N.~P. Warner, ``{Black holes, black rings and their microstates},''
  {\em Lect.Notes Phys.} {\bf 755} (2008) 1--92,
\href{http://www.arXiv.org/abs/hep-th/0701216}{{\tt hep-th/0701216}}.

\bibitem{Skenderis:2008qn}
K.~Skenderis and M.~Taylor, ``{The fuzzball proposal for black holes},'' {\em
  Phys.Rept.} {\bf 467} (2008) 117--171,
\href{http://www.arXiv.org/abs/0804.0552}{{\tt 0804.0552}}.

\bibitem{Mathur:2008nj}
S.~D. Mathur, ``{Fuzzballs and the information paradox: A Summary and
  conjectures},''
\href{http://www.arXiv.org/abs/0810.4525}{{\tt 0810.4525}}.

\bibitem{Almheiri:2012rt}
A.~Almheiri, D.~Marolf, J.~Polchinski, and J.~Sully, ``{Black Holes:
  Complementarity or Firewalls?},'' {\em JHEP} {\bf 1302} (2013) 062,
\href{http://www.arXiv.org/abs/1207.3123}{{\tt 1207.3123}};
 cf. S. L. Braunstein, ``{Black hole entropy as entropy of
entanglement, or it's curtains for the equivalence principle},"
\href{http://www.arXiv.org/abs/0907.1190}{{\tt 0907.1190}} published 
as S.~L.~Braunstein, S.~Pirandola and K.~Zyczkowski, ``{Better Late than Never: Information
Retrieval from Black Holes}," {\em Phys.Rev.Lett.} {\bf 110} (2013) 101301
for a similar prediction from different assumptions.

\bibitem{Almheiri:2013hfa}
A.~Almheiri, D.~Marolf, J.~Polchinski, D.~Stanford, and J.~Sully, ``{An
  Apologia for Firewalls},'' {\em JHEP} {\bf 1309} (2013) 018,
\href{http://www.arXiv.org/abs/1304.6483}{{\tt 1304.6483}}.

\bibitem{Marolf:2013dba}
D.~Marolf and J.~Polchinski, ``{Gauge/Gravity Duality and the Black Hole
  Interior},'' {\em Phys.Rev.Lett.} {\bf 111} (2013) 171301,
\href{http://www.arXiv.org/abs/1307.4706}{{\tt 1307.4706}}.

\bibitem{Carlip:2000nv}
S.~Carlip, ``{Logarithmic corrections to black hole entropy from the Cardy
  formula},'' {\em Class. Quant. Grav.} {\bf 17} (2000) 4175--4186,
\href{http://www.arXiv.org/abs/gr-qc/0005017}{{\tt gr-qc/0005017}}.

\bibitem{Sen:2012dw}
A.~Sen, ``{Logarithmic Corrections to Schwarzschild and Other Non-extremal
  Black Hole Entropy in Different Dimensions},'' {\em JHEP} {\bf 1304} (2013)
  156,
\href{http://www.arXiv.org/abs/1205.0971}{{\tt 1205.0971}}.

\bibitem{Kaul:1998xv}
R.~K. Kaul and P.~Majumdar, ``Quantum black hole entropy,'' {\em Phys. Lett.}
  {\bf B439} (1998) 267--270,
\href{http://www.arXiv.org/abs/gr-qc/9801080}{{\tt gr-qc/9801080}}.

\bibitem{Kaul:2000kf}
R.~K. Kaul and P.~Majumdar, ``{Logarithmic correction to the Bekenstein-Hawking
  entropy},'' {\em Phys. Rev. Lett.} {\bf 84} (2000) 5255--5257,
\href{http://www.arXiv.org/abs/gr-qc/0002040}{{\tt gr-qc/0002040}}.

\bibitem{IBM}
T.~Watson, ``{I} think there is a world market for maybe five computers.''
  [president of IBM], 1943.

\bibitem{Wheeler}
J.~Wheeler, ``{E}verything is information'' or ``{I}t from bit'' was a slogan
  that replaced {W}heeler's earlier belief ``everything is geometry''.
  Indeed, taking geometry too seriously can be misleading in the context of
  quantum gravity. 1990.

\bibitem{Maldacena:2013xja}
J.~Maldacena and L.~Susskind, ``{Cool horizons for entangled black holes},''
{\em Fortsch. Phys. }{\bf 61} (2013) 781-811
\href{http://www.arXiv.org/abs/1306.0533}{{\tt 1306.0533}}.



\end{thebibliography}

\providecommand{\href}[2]{#2}\begingroup\raggedright\endgroup

\end{document}